\pdfoutput=0
\documentclass[journal,11pt,draftclsnofoot,onecolumn]{IEEEtran}

\usepackage{graphicx}
\DeclareGraphicsExtensions{.eps}

\usepackage{subfigure, epsfig}

\newfont{\bbb}{msbm10 scaled 500}

\newfont{\bb}{msbm10 scaled 1100}

\newcommand{\Dm}{{\bf D}}

\newcommand{\Hm}{{\bf H}}

\newcommand{\Qm}{{\bf Q}}

\newcommand{\Sm}{{\bf S}}

\newcommand{\Um}{{\bf U}}

\newcommand{\Vm}{{\bf V}}

\newcommand{\diag}{{\hbox{diag}}}

\usepackage{amsfonts}
\usepackage{amssymb}
\usepackage{amsmath}
\usepackage{float}
\usepackage{color}
\usepackage{amscd}                 

\newcommand{\mb}{\mathbf}
\newcommand{\ul}{\underline}

\newcommand{\ds}{\displaystyle}

\newtheorem{Lemma}{Lemma}

\newtheorem{theorem}{Theorem}[section]

\newtheorem{conjecture}[theorem]{Conjecture}
\newtheorem{proposition}[theorem]{Proposition}

\begin{document}
\title{An Energy-Efficient Framework for the Analysis of MIMO Slow Fading Channels}
 \author{Vineeth S. Varma, Samson Lasaulce, Merouane Debbah,~and Salah Eddine Elayoubi,%
\thanks{V. S. Varma and S. E. Elayoubi are with the Orange Labs, 92130 Issy Les Moulineaux, France (e-mail: vineethsvarma@gmail.com; salaheddine.elayoubi@orange.com).}
\thanks{S. Lasaulce and M. Debbah are with SUPELEC, 91192 Gif-sur-Yvette, France (e-mail: Samson.LASAULCE@lss.supelec.fr; merouane.debbah@supelec.fr).}
}%

\markboth{Journal of IEEE, Trans in Signal Processing}%
{Trans}%
\maketitle

\begin{abstract}
In this work, a new energy-efficiency performance metric is proposed for MIMO (multiple input multiple output) point-to-point systems. In contrast with related works on energy-efficiency, this metric translates the effects of using finite blocks for transmitting, using channel estimates at the transmitter and receiver, and considering the total power consumed by the transmitter instead of the radiated power only. The main objective pursued is to choose the best pre-coding matrix used at the transmitter in the following two scenarios~: 1) the one where imperfect channel state information (CSI) is available at the transmitter and receiver~; 2) the one where no CSI is available at the transmitter. In both scenarios, the problem of optimally tuning the total used power is shown to be non-trivial. In scenario 2), the optimal fraction of training time can be characterized by a simple equation. These results and others provided in the paper, along with the provided numerical analysis, show that the present work can therefore be used as a good basis for studying power control and resource allocation in energy-efficient multiuser networks.

\end{abstract}


%

\section{Introduction}\label{sec:introduction}

Over the past two decades, designing energy-efficient communication terminals has become an important issue. This is not surprising for terminals which have to be autonomous as far as energy is concerned, such as cellular phones, unplugged laptops, wireless sensors, and mobile robots. More surprisingly, energy consumption has also become a critical issue for the fixed infrastructure of wireless networks. For instance, Vodafone's global energy consumption for 2007-2008 was about 3000 GWh \cite{genref1}, which corresponds to emitting 1.45 million tons of CO2 and represents a monetary cost of a few hundred million Euros. This context explains, in part, why concepts like ``green communications'' have emerged as seen from \cite{palicot,genref2} and \cite{genref3}. Using large multiple antennas, virtual multiple input multiple output (MIMO) systems, and small cells is envisioned to be one way of contributing to reducing energy consumption drastically. The work reported in this paper concerns point-to-point MIMO systems in which communication links evolve in a quasi-static manner, these channels are referred to as MIMO slow fading channels. The performance metric considered for measuring energy-efficiency of a MIMO communication corresponds to a trade-off between the net transmission rate (transmission benefit) and the consumed power (transmission cost).

The ultimate goal pursued in this paper is a relatively important problem in signal processing for communications. It consists of tuning the covariance matrix of the transmitted signal (called the pre-coding matrix) optimally. But, in contrast with the vast literature initiated by \cite{telatar} in which the transmission rate is of prime interest, the present paper aims at optimizing the pre-coding matrix in the sense of energy-efficiency as stated in \cite{veronica2}. Interestingly, in \cite{veronica2} the authors bridge a gap between the pioneering work by Verd\'{u} on the capacity per unit cost for static channels \cite{verdu} and the more pragmatic definition of energy-efficiency proposed by \cite{goodman} for quasi-static single input single output (SISO) channels. Indeed, in \cite{veronica2}, energy-efficiency is defined as the ratio of the probability that the channel mutual information is greater than a given threshold to the used transmit power. Assuming perfect channel state information at the receiver (CSIR) and the knowledge of the channel distribution at the transmitter, the pre-coding matrix is then optimized for several special cases. While \cite{veronica2} provides interesting insights into how to allocate and control power at the transmitter, a critical issue is left unanswered; to what extent do the conclusions of \cite{veronica2} hold in more practical scenarios such as those involving imperfect CSI? Answering this question was one of the motivations for the work reported here. Below, the main differences between the approach used in this work and several existing relevant works are reviewed.

In the proposed approach, the goal pursued is to maximize the number of information bits transmitted successfully per Joule consumed at the transmitter. This is different from the most conventional approach which consists in minimizing the transmit power under a transmission rate constraint: \cite{cui} perfectly represents this body of literature. In the latter and related works, efficiency is not the main motivation. \cite{tse} provides a good motivation as to how energy-efficiency can be more relevant than minimizing power under a rate constraint. Indeed, in a communication system without delay constraints, rate constraints are generally irrelevant whereas the way energy is used to transmit the (sporadic) packets is of prime interest. Rather, our approach follows the original works on energy-efficiency which includes \cite{shah,goodman,saraydar,similar,buzzi}. The current state of the art indicates that, since \cite{veronica2}, there have been no works where the MIMO case is treated by exploiting the cumulative distribution of the channel mutual information (i.e., the outage probability) at the numerator of the performance metric. As explained below, our analysis goes much further than \cite{veronica2} by considering effects such as channel estimation error effects. In the latter respect, several works address the issue of power allocation for outage probability minimization \cite{yoo-goldsmith-2007,giuseppe-it-2010,cesar-08} under imperfect channel state information. The latter will serve as a basis for the analysis conducted in the present paper. At this point, it is possible to state the contributions of the present work.

In comparison to \cite{veronica2}, which is the closest related work, the main contributions of the paper can be summarized as follows:
\begin{itemize}
  \item one of the scenarios under investigation concerns he case where CSI is also available at the transmitter (only the case with CSIR and CSI distribution at the transmitter is studied in \cite{veronica2}).

  \item The assumption of perfect CSI is relaxed. In Sec. III, it is assumed that only imperfect CSIT and imperfect CSIR is available. Sec. IV considers the case with no CSIT and imperfect CSIR. In particular, this leads us to the problem of tuning the fraction of training time optimally. Exploiting existing works for the transmission rate analysis \cite{mimo} and \cite{mimot}, it is shown that this problem can also be treated for energy-efficiency.

  \item The realistic assumption of finite block length is made. This is particularly relevant, since block finiteness is also a source of outage and therefore impacts energy-efficiency. Note that recent works on transmission under the finite length regime such as \cite{polynaskiy} provide a powerful theoretical background for possible extensions of this paper.

 \item Instead of considering the radiated power only for the cost of transmitting, the total power consumed by the transmitter is accounted for. Based on works such as \cite{richter}, an affine relation between the two is assumed. Although more advanced models can be assumed, this change is sufficient to show that the behavior of energy-efficiency is also modified.

\end{itemize}

The paper is therefore structured as follows. Sec. II describes the proposed framework to tackle the aforementioned issues. Sec. III and IV treat the case with and without CSIT respectively. They are followed by a section dedicated to numerical results (Sec. V) whereas Sec. \ref{sec:conclusion} concludes the paper with the main messages of this paper and some relevant extensions.

\section{System model}\label{sec:system-model}

A point-to-point multiple input and multiple output communication unit is studied in this work. In this paper, the dimensionality of the input and output is given by the numbers of antennas but the analysis holds for other scenarios such as virtual MIMO systems \cite{vmimo}. If the total transmit power is given as $P$, the average SNR is given by~:
\begin{equation}\label{eq:def-snr}
\rho = \frac{P}{\sigma^2}
\end{equation}
where $\sigma^2$ is the reception noise variance.The signal at the receiver is modeled by~:
\begin{equation}
\label{eq:system-model-mimo}
 \ul{y}=  \sqrt{\frac{\rho}{M}}  \mb{H}  \ul{s}
+ \ul{z}
\end{equation}
where $\mb{H}$ is the $N \times M$ channel transfer matrix and $M$ (resp. $N$) the number of transmit (resp. receive) antennas. The entries of $\mb{H}$ are i.i.d. zero-mean unit-variance complex Gaussian random variables. The vector $\ul{s}$ is the $M$-dimensional column vector of transmitted symbols follows a complex normal distribution, and $\ul{z}$ is an $N$-dimensional complex white Gaussian noise distributed as $\mathcal{N}(\ul{0}, \mb{I})$. Denoting by $\mb{Q} = \mathbb{E}[\ul{s}\ul{s}^H]$ the input covariance matrix (called the pre-coding matrix), which satisfies
\begin{equation}
\label{eq:pre-coding}  \frac{1}{M} \mathrm{Tr}(\mb{Q}) =  1
\end{equation}
where $\mathrm{Tr}$ stands for the trace operator. The power constraint is expressed as~:
\begin{equation}
\label{eq:power-contraint}  P \leq P_{\max}
\end{equation}
where $P_{\max}$ is the maximum available power at the transmitter. 

The channel matrix $\mb{H}$ is assumed to evolve in a quasi-static manner~: the channel is constant for some time interval, after which it changes to an independent value that it holds for the next interval \cite{mimo}. This model is appropriate for the slow-fading case where the time with which $\Hm$ changes is much larger than the symbol duration.

\subsection{Defining the energy efficiency metric}

In this section, we introduce and justify the proposed definition of energy-efficiency of a communication system with multiple input and output antennas, and experiences slow fading.

In \cite{goodman}, the authors study multiple access channels with SISO links and use the properties of the energy efficiency function defined as $\frac{f(\rho)}{P}$ to establish a relation between the channel state (channel complex gain) ($h$) and the optimal power ($P^*$). This can be written as:
\begin{equation}
P^*=\frac{\mathrm{SNR}^* \sigma^2}{|h|^2}
\end{equation}
where $\mathrm{SNR}^*$ is the optimal SNR for any channel state and (when $f$ is a sigmoidal/S-shaped function, i.e, it is initially convex and after some point becomes concave) is the unique strictly positive
solution of
\begin{equation}
x f'(x) - f(x)=0
\end{equation}
where $1-f(.)$ is the outage probability. Formulating this problem in the case of MIMO channels is non-trivial as there is a problem of choosing the total transmit power as well as the power allocation.

When the same (imperfect) CSI is available at the transmitter and receiver, by estimating the channel for $t$ time, and sending the information to the transmitter for $t_f$ time, the energy-efficiency $\nu_T$ is defined as:
\begin{equation}\label{eq:def-ee-csi}
\nu_T(P,\Qm,\hat{\Hm}) = \frac{R  \left(1 - \frac{t+t_f}{T}\right) F_{L} \left[ I_{\mathrm{ICSITR}}(P,\Qm,\hat{\Hm})-\frac{R}{R_0} \right]  } {a P +b}
\end{equation}

where $R$ is the transmission rate in bit/s, $T$ is the block duration in s, $R_0$ is a parameter which has unit Hz (e.g., the system bandwidth), and $a>0$, $b\geq0$ are parameters to relate the transmitter radiated power to its total consumed power~; we define $\xi = \frac{R}{R_0}$ as the spectral efficiency. $I_{\mathrm{ICSITR}}(P,\Qm,\hat{\Hm})$ denotes the mutual information with imperfect CSITR (the receiver also has the exact same CSI as the transmitter). This form of the energy-efficiency is inspired from early definitions provided in works like \cite{goodman}, and studies the gain in data rate with respect to the cost which is the power consumed. The numerator represents the benefit associated with transmitting namely, the net transmission rate (called the goodput in \cite{goodput}) of the communication and is measured in bit/s. The goodput comprises a term $1 - \frac{t+t_f}{T}$ which represents the loss in terms of information rate due to the presence of a training and feedback mechanism (for duration $t$ seconds and $t_f$ seconds resp. in a $T$s long block) \footnote{In this case, we assume that the feedback mechanism is sufficient to result in perfect knowledge of $\hat{\Hm}$ at the transmitter. This is done because, assuming a different imperfect CSI at the transmitter from the receiver creates too much complexity and this problem is beyond the scope of a single paper.}. The denominator of (\ref{eq:def-ee-csi}) represents the cost of transmission in terms of power. The proposed form for the denominator of (\ref{eq:def-ee-csi}) is inspired from \cite{richter} where the authors propose to relate the average power consumption of a transmitter (base stations in their case),to the average radiated or radio-frequency power by an affine model.

The term $F_L(.)$ represents the transmission success probability. The quantity $F_L(.)$ gives the probability that the ``information'' denoted by $\hat{I}$ as defined in \cite{Buckingham-08}) is greater than or equal to the coding rate ($\xi$), i.e., it is the complementary cumulative distribution function of the information $\hat{I}$, $\mathrm{Prob}( \hat{I} \geq \xi)$. Formally, $\hat{I}$ is defined as $\hat{I}=\log \frac{\mathrm{PDF}_{X,Y}(x,y)}{\mathrm{PDF}_{X}(x)\mathrm{PDF}_{Y}(y)}$, where PDF$_{X,Y}$ and PDF$_{X}$ represents the joint and marginal probability distribution functions, $x$ and $y$ are samples of the process $X$ and $Y$, which in this case represent the transmitted and received signals. The average mutual information $I = E(\hat{I})$ is used to calculate this probability and $F_L(.)$ depends on the difference between $I$ and $\xi$. $F_L(.)$ can be verified to be sigmoidal (this is the cumulative probability distribution function of a variable with a single peaked probability distribution function) and $F_L(0)=0.5$ (If $\xi=I$, $F_L(.)$ is the probability that a random variable is equal to or larger than its mean). When CSIT is available, it is possible to ensure that the data transmission rate is just below the channel capacity. If this is done, then there is no possibility of outage when the block length is infinite \cite{shannon}. However, in most practical cases, the block length is finite and this creates an outage effect which depends on the block length $L$ \cite{Buckingham-08}.

The bounds on $F_L$ can be expressed as $F_L(I_{\mathrm{ICSITR}}(0,0,\Hm)-\xi)=0$ (no reliable communication when transmit power is zero) and as $F_L \to 1$ when $P \to \infty$. This proposed form for this function, $F_L(I_{\mathrm{ICSITR}}(P,\Qm,\hat{\Hm})-\xi)$, is supported by works like \cite{Buckingham-08} and \cite{hoydis-2012}. An approximation for this function based on the automatic repeat request protocol \cite{arq} is $F_L(x)=Q_{func}(-Tx)$, where $Q_{func}$ is the tail probability of the standard normal distribution.

Therefore, in the presence of CSI at the transmitter, outage occurs even when the mutual information is more than the targeted rate due to the noise and finite code-lengths. In this scenario, the energy-efficiency is maximized when the parameters $\Qm$ and $P$ are optimized.

In the absence of CSI at the transmitter, the earlier definition of energy efficiency is not suitable since $\Hm$ is random, $\nu_{T}$ is also a random quantity. Additionally, in this case, it is impossible to know if the data transmission rate is lower than the instantaneous channel capacity as the channel varies from block to block. Therefore, in this case, the source of outage is primarily the variation of the channel \cite{ozarow-1994}, and using (\ref{eq:def-ee-csi}) directly is not suitable. As the channel information is unavailable at the transmitter, define $\mb{Q} = \frac{\mb{I}_M}{M}$, meaning that the transmit power is allocated uniformly over the transmit antennas; in Sec. \ref{sec:subsec-opt-M}, we will comment more on this assumption. Under this assumption, the average energy-efficiency can be calculated as the expectation of the instantaneous energy-efficiency over all possible channel realizations. This can be rewritten as:
\begin{equation}
\nu_R(P,t) = \frac{R \left(1 - \frac{t}{T}\right)
 \mathbb{E}_{\Hm}\left( F_L\left[ I_{\mathrm{ICSIR}}(P,\Qm,\hat{\Hm})-\frac{R}{R_0} \right] \right)  }{a P +b}.
\end{equation}
For large $L$, it has been shown in \cite{ozarow-1994} (and later used in other works like \cite{veronica2}) that the above equation can be well approximated to~:
\begin{equation}\label{eq:def-ee-nocsit}
\nu_R(P,t) = \frac{R \left(1 - \frac{t}{T}\right)\mathrm{Pr}_{\Hm} \left[ I_{\mathrm{ICSIR}}(P,t, \hat{\Hm}) \geq \xi   \right]  }{a P +b}
\end{equation}
where $\mathrm{Pr}_{\Hm}$ represents the probability evaluated over the realizations of the random variable $\Hm$. Here, $I_{ICSIR}$ represents the mutual information of the channel with imperfect CSI at the receiver. Let us comment on this definition of energy efficiency. This definition is similar to the earlier definition in all most ways. Here the parameter $t$, represents the length of the training sequence used to learn the channel at the receiver\footnote{In this case, the optimization is done over $P$ and $t$ assuming imperfect CSI at the receiver. A parameter here not explicitly stated, but indicated nevertheless, is $M$ due to the number of transmit antennas affecting the effectiveness of training}. The major difference here is that the expression for the success rate is the probability that the associated mutual information is above a certain threshold. This definition of the outage is shown to be appropriate and compatible with the earlier definition when only statistical knowledge of the channel is available \cite{ozarow-1994}.

Although very simple, these models allow one, in particular, to study two regimes of interest.
\begin{itemize}
  \item The regime where $\frac{b}{a}$ is small allows one to study not only communication systems where the power consumed by the transmitter is determined by the radiated power but also those which have to been green in terms of electromagnetic pollution or due to wireless signal restrictions (see e.g., \cite{wifir}).

  \item The regime where $\frac{b}{a}$ is large allows one to study not only communication systems where the consumed power is almost independent of the radiated power but also those where the performance criterion is the goodput.
\end{itemize}
Note that when $b=0$, $t \rightarrow +\infty$, $T \rightarrow +\infty$, and $\frac{t}{T}
\rightarrow 0$ equation (\ref{eq:def-ee-nocsit}) boils down to the performance
metric investigated in \cite{veronica2}.

\subsection{Modeling channel estimation noise}

Each transmitted block of data is assumed to comprise a training sequence in order for the receiver to be able to estimate the channel; the training sequence length in symbols is denoted by $t_s$ and the block length in symbols by $T_s$. Continuous counterparts of the latter quantities are defined by $t= t_s  S_d$ and $T = T_s  S_d$, where $S_d$ is the symbol duration in seconds. In the training phase, all $M$ transmitting antennas broadcast orthogonal sequences of known pilot/training symbols of equal power on all antennas. The receiver estimates the channel, based on the observation of the training sequence, as $\widehat{\Hm}$ and the error in estimation is given as $\Delta \Hm = \Hm - \widehat{\Hm}$. Concerning the number of observations needed to estimate the channel, note that typical channel estimators generally require at least as many measurements as unknowns \cite{mimot}, that is to say $N t_s \geq  N M$ or more simply
\begin{equation}\label{eq:condition-ts-M}
t_s \geq  M.
\end{equation}
The channel estimate normalized to unit variance is denoted by $\widetilde{\Hm}$. From \cite{mimot} we know that the mutual information is the lowest when the estimation noise is Gaussian. Taking the worst case noise, it has been shown in \cite{mimo} that the following observation equation
\begin{equation}\label{eq:cmodel2n}
\widetilde{\ul{y}}  =   \sqrt{\frac{\rho_{\mathrm{eff}}(\rho,t)}{M}}\widetilde{\Hm}
\ul{s} + \widetilde{\ul{z}}
\end{equation}
perfectly translates the loss in terms of mutual information\footnote{It is implicitly assumed that the mutual information is taken between the system input and output; this quantity is known to be very relevant to characterize the transmission quality of a communication system (see e.g. \cite{cover} for a definition).} due to channel estimation provided that the effective SNR $\rho_{\mathrm{eff}}(\rho,t)$ and equivalent observation noise $\widetilde{\ul{z}}$ are defined properly namely, 
\begin{equation}
\label{eq:rhofun}
\left\{
\begin{array}{ccc}
\widetilde{\ul{z}} & = & \sqrt{\frac{\rho}{M}} \Delta \Hm \ul{s} +\ul{z}\\
\rho_{\mathrm{eff}}(\rho,t) & = & \frac{\frac{t}{M S_d}\rho^2}{1+\rho + \rho \frac{t}{M S_d}}
\end{array}
.
\right.
\end{equation}
As the worst case scenario for the estimation noise is assumed, all formulas derived in the following sections give lower bounds on the mutual information and success rates. Note that the lower bound is tight (in fact, the lower bound is equal to the actual mutual information) when the estimation noise is Gaussian which is true in practical cases of channel estimation.  The effectiveness of this model will not be discussed here but has been confirmed in many other works of practical interest (see e.g., \cite{samson}). Note that the above equation can be utilized for the cases of imperfect CSITR and CSIR as well as the case of imperfect CSIR with no CSITR. This is because in both cases, the outage is determined by calculating the mutual information $I_{\mathrm{ICSITR}}$ or $I_{\mathrm{ICSIR}}$ respectively.

\section{Optimizing energy-efficiency with imperfect CSITR available}
\label{sec:optee-csit}

When perfect CSITR or CSIR is available, the mutual information of a MIMO system, with a pre-coding scheme $\Qm$ and channel matrix $\Hm$ can be expressed as:
\begin{equation}\label{eq:def-cappcsit}
I_{\mathrm{CSITR}}(P,\Qm,\Hm) = \log \left| \mb{I}_M + \frac{P}{M\sigma^2} \Hm \Qm
 \Hm^H \right|
\end{equation}
The notation $|\mb{A}|$ denotes the determinant of the (square) matrix $\mb{A}$.
With imperfect CSIT, which is exactly the same as the CSIR (i.e., both the transmitter and the receiver have the same channel estimate $\hat{\Hm}$), a lower bound on the mutual information can be found from several works like \cite{yoo-goldsmith-2007,cesar-08} etc. This lower bound for $I_{\mathrm{ICSITR}}$ is used, which is expressed as:
\begin{equation}\label{eq:def-capipcsit-1}
I_{\mathrm{ICSITR}}(P,\Qm,\hat{\Hm}) = \log \left| \mb{I}_M + \hat{\Hm} \frac{P}{M\sigma^2(1+\rho \sigma^2_E) } \Qm \hat{\Hm}^H \right|
\end{equation}
where $\hat{\Hm}$ is the estimated channel and $1-\sigma^2_E$ is the variance of $\hat{\Hm}$. Considering the block fading channel model, from \cite{yoo-goldsmith-2007} and \cite{mimot} we conclude that $\sigma^2_E=\frac{1}{1+\rho \frac{t}{M}}$. Simplifying :
\begin{equation}\label{eq:def-capipcsit}
I_{\mathrm{ICSITR}}(P,\Qm,\widehat{\Hm}) = \log \left| \mb{I}_M + \frac{\rho_{\text{eff}}}{M} \widehat{\Hm} \Qm
 \widehat{\Hm}^H \right|.
\end{equation}
Having defined the mutual information to be used for (\ref{eq:def-ee-csi}), we proceed with optimizing $\nu_T$.

\subsection{Optimizing the pre-coding matrix $\Qm$}

Studying (\ref{eq:def-ee-csi}) and (\ref{eq:def-capipcsit}), we see that varying the power allocation (or the corresponding pre-coding matrix) $\Qm$, affects only the success rate $F_L(.)$ and the total power $P$ is the only term that is present outside $F_L(.)$. As $F_L(.)$ is known to be an increasing function, if the total power is a constant, optimizing the energy efficiency $\nu_T$ amounts to simply maximizing the mutual information $I_{\mathrm{ICSITR}}(P,\Qm,\hat{\Hm})$. This is a well documented problem and it gives a ``water-filling'' type of solution \cite{love-2008}. Rewriting (\ref{eq:def-cappcsit}) as
\begin{equation}\label{eq:def-cappcsitmod}
I_{\mathrm{ICSITR}}(P,\Qm,\widehat{\Hm}) = \log \left| \mb{I}_M + \frac{\rho_{\mathrm{eff}}}{M} \Dm \Sm
 \Dm^H \right|
\end{equation}
where the optimal covariance matrix $\Qm=\Vm\Sm\Vm^H$ is achieved through the singular value decomposition of the channel matrix $\widehat{\Hm}=\Um\Dm\Vm^H$ and an optimal diagonal covariance matrix $\Sm=\diag[s_1, \dots, s_{\min(M,N)},0, \dots,0]$. The water-filling algorithm can be performed by solving:
\begin{align} \label{eq:wf}
& s_i  =  \left( \mu - \frac{1}{\rho \|d_i\|^2} \right)^{+} ,\:\mathrm{ for }\: i=1,2,\cdots,{\min(M,N)}
\end{align}
where $d_i$ are the diagonal elements of $\Dm$ and $\mu$ is selected such that $\Sigma_{i=1}^{\min(M,N)} s_i =M$. Here $(x)^+ = \max(0,x)$, this implies that $s_i$ can never be negative. The actual number of non-vanishing entries in $\Sm$ depends on the values of $d_i$ as well $\rho$ (and thus $P$). Examining (\ref{eq:wf}), we can see that when $\rho \to 0$, the water-filling algorithm will lead to choosing $s_j=M$ and $s_i=0$ for all $i \neq j$, where $j$ is chosen such that $d_j=\max(d_i)$ (beamforming). Similarly for $\rho \to \infty $, $s_i=\frac{M}{\min(M,N)}$ (uniform power allocation).
%

\subsection{Determining the optimal total power}

$\Qm$ has been optimized in the previous section. From (\ref{eq:def-ee-csi}), we see that the parameters that can be optimized in order to maximize the energy efficiency are $\Qm$ and $P$. Therefore, in this section, we try to optimize $P$, the total power. Note that for every different $P$, the optimal power allocation $\Qm$ changes according to (\ref{eq:wf}) as $\rho$ is directly proportional to $P$. Therefore optimizing this parameter is not a trivial exercise. Practically, $P$ represents the total radio power, that is, the total power transmitted by the antennas. This power determines the total consumed power $b+aP$, of base stations or mobile terminals and so, optimizing this power is of great importance.

In this section, a theorem on the properties of $\nu_T(P,\Qm_{WF(P)},\widehat{\Hm})$ is provided, where $\Qm_{WF(P)}$ is the power allocation obtained by using the water-filling algorithm and iteratively solving (\ref{eq:wf}) with power $P$. This procedure is said to be ``iterative" because, after solving equation \ref{eq:wf}, if any $s_j<0$, then we set $s_j=0$ and the equation is resolved until the all solutions are positive. For optimization, desirable properties on $\nu_T(P,\Qm_{WF(P)},\widehat{\Hm})$ are differentiability, quasi-concavity and the existence of a maximum. The following theorem states that these properties are in fact satisfied by $\nu_T$.

\begin{theorem}
\label{th:qcwf}
The energy-efficiency function $\nu_T(P,\Qm_{WF(P)},\widehat{\Hm}) $ is quasi-concave with respect to $P$ and has a unique maximum $\nu_T(P^*,\Qm_{WF(P^*)},\widehat{\Hm})$, where $P^*$ satisfies the following equation~:
\begin{eqnarray}\label{eq:optp-csit}
&\frac{ \partial F_L[I_{\mathrm{ICSITR}}(P^*,\Qm_{WF(P^*)},\widehat{\Hm})-\xi]}{\partial P}\left(P^*+\frac{b}{a}\right) &\\ \nonumber
&-F_L[I_{\mathrm{ICSITR}}(P^*,\Qm_{WF(P^*)},\widehat{\Hm})-\xi]&=0
\end{eqnarray}where $\frac{\partial}{\partial P} $ is the partial derivative.
\end{theorem}

The proof of this theorem can be found in Appendix \ref{sec:proofqcwf}. From the above theorem and equation, we can conclude that the optimal transmit power for imperfect CSITR depends on several factors like
\begin{itemize}
\item the channel estimate $\widehat{\Hm}$,
\item the target spectral efficiency $\xi$,
\item the ratio of the constant power consumption to the radio-frequency (RF) power efficiency $\frac{b}{a}$,
\item the channel training time $t$ and
\item the noise level $\sigma^2$.
\end{itemize}
Note that in this model, we always assume the CSI at the transmitter to be exactly identical to CSI at the receiver. Because of this, we take the feedback mechanism to be perfect and take a constant time $t_f$. Although in practice, $t_f$ plays a role in determining the efficiency and the optimal power, in our model $t_f$ is a constant and does not appear in the equation for $P^*$. In our numerical results we focus on the impact of $\widehat{\Hm}$, $\xi$ and $\frac{b}{a}$ on $P^*$ and $\nu^*$. The impact of $t$ is not considered for this case, but is instead studied where we have no CSITR and imperfect CSIR, this choice helps in making the results presented easier to interpret and understand.

\subsection{An illustrative special case~: SISO channels}

A study on energy-efficiency in SISO systems have been studied in many works like \cite{goodman} and \cite{verdu}. However, the approach used in this paper is quite novel even for the SISO case and presents some interesting insights that have not been presented before. For the case of SISO, the pre-coding matrix is a scalar and $\Qm=1$. The optimal power can be determined by solving (\ref{eq:optp-csit}). For a SISO system with perfect CSITR and CSIR, $F_L$ can be expressed as
\begin{eqnarray}
&F_L[I_{\mathrm{ICSITR}}(P^*,\Qm_{WF(P^*)},\widehat{\Hm})-\xi]= &\nonumber \\
& Q_{\mathrm{function}}\left(L (1+\|h\|^2 \rho) \frac{\xi-\log(1+\|h\|^2 \rho )}{ \|h\|^2 \rho}\right)&
\end{eqnarray}
from \cite{Buckingham-08}. Using this expression, we can find $P^*$ maximizing $\nu_T$.

In the case of high SNR (and high $\xi$), a solution to this problem can be found as
\begin{equation}
\lim_{\rho \to \infty} F_L(P,1,h)= Q_{func}\left(  L [\xi-\log(1+\|h\|^2 \rho] )\right).
\end{equation}
Solving (\ref{eq:optp-csit})
\begin{eqnarray}
&\frac{-L \|h\|^2  }{ \sqrt{\pi}(1+\|h\|^2 \rho^*)} \exp\left( - L^2 \left[ \xi-\log(1+\|h\|^2 \rho^* )\right]^2  \right) \left(\rho^*+\frac{b}{a\sigma^2}\right) &\nonumber\\ 
&- Q_{func}\left( L [\xi-\log(1+\|h\|^2 \rho^*] )\right) =0&.
\end{eqnarray}
From the above equation it can be deduced that if $b=0$, for large $\xi$, $\log(1+\rho^*) \approx \xi$. While for low SNR, $\lim_{\rho \to 0} F_L(P,1,h) = Q_{func}\left( L \frac{\xi-\log(1+\|h\|^2 \rho )}{\|h\|^2 \rho}\right)$ and so, if $b=0$,
\begin{eqnarray}
&\frac{1 }{ \sqrt{\pi}}\left[  L + L\frac{\xi-|h\|^2 \rho^*}{  \|h\|^2 \rho^*}  \right] \exp\left( \frac{-1}{2}\left[ L\frac{\xi-\|h\|^2 \rho^* }{\|h\|^2 \rho^*}\right]^2 \right)& \\ \nonumber- &Q_{func}\left( L \frac{\xi-\|h\|^2 \rho^* }{\|h\|^2 \rho^*}\right) & = 0
\end{eqnarray}
Substitute $x= L \frac{\xi-\|h\|^2 \rho^* }{\|h\|^2 \rho^*}$ and we have
\begin{equation}
\frac{1 }{ \sqrt{\pi}}\left[  L + x  \right] \exp\left( \frac{-1}{2} x^2 \right) - Q_{func}\left( x \right)=0.
\end{equation}

As seen from the above equation, the value of $x$ depends only on $L$ the block length. For example if $L=10$, we get $x \approx -1.3$. So, $\rho^* = 1.14 \frac{\xi}{\|h\|^2}$. Whereas if $L=100$ we get $\rho^* = 1.02  \frac{\xi}{\|h\|^2}$. Note that these calculations are true only for $\xi \to 0$ so that $\rho \to 0$ is satisfied.

The above equations signify that for finite block lengths, the energy efficiency at $\xi \to 0$ is lower than the value calculated in \cite{verdu} (of course, a direct comparison does not make sense as in \cite{verdu}, infinite block lengths are assumed). This suggests that a non-zero value of $\xi$ might optimize the energy efficiency. This value is evaluated in our numerical section and we find that the energy efficiency is optimized at a non-zero power.

\subsection{Special Case: Infinite code-length and perfect CSITR}

When a very large block is used then the achievable rate approaches the mutual information \cite{shannon}, i.e $\lim_{L \to \infty , I_{\mathrm{CSITR}}-\xi \to 0^+} F_L(I_{\mathrm{CSITR}}-\xi) =1$. Therefore in this limit, we can now simplify (\ref{eq:def-ee-csi}) to:
\begin{equation}
\nu_T(P,\Qm,\widehat{\Hm}) = \frac{R_0  \left(1 - \frac{t+t_f}{T}\right)  I_{\mathrm{CSITR}}(P,\Qm,\widehat{\Hm}) }  {a P +b}.
\end{equation}
This is done because we replace $\xi$ with $I_{\mathrm{CSITR}}$ to maximize efficiency as $F_L$ is $0$ when $I_{\mathrm{CSITR}} < \xi$, and choosing $\xi \to I_{\mathrm{CSITR}}$ maximizes efficiency. Water-filling optimizes the efficiency in this situation as well, and so we use $\Qm=\Qm_{WF(P)}$. It can be easily verified that for $b \to 0$: $\nu_T$ is maximized for $P \to 0$. And in this case, water-filling also implies that only the antenna with the best channel is used to transmit. Interestingly, when in the domain of finite code-lengths, our simulations indicate that there is a non-zero rate and power that optimizes the energy-efficiency function.

For general $b$, $\nu_T$ is optimized for $P^*$ satisfying:
\begin{equation}\label{eq:ee-infcl}
 \frac{\partial I_{\mathrm{CSITR}}(P,\Qm_{WF(P)},\widehat{\Hm}) }{\partial P}  (a P +b) - I_{\mathrm{CSITR}}(P,\Qm,\widehat{\Hm}) =0.
\end{equation}

The above equation admits a unique maximum because $I_{\mathrm{CSITR}}(P,\Qm_{WF(P)},\widehat{\Hm})$ is a concave function of $P$ (can be seen from Appendix \ref{sec:proofqcwf}) and is mathematically appealing. For $\lim_{b \to 0} P^*=0$ and as $\frac{b}{a}$ increases, $P^*$ also increases. A special case of this, with $b=0$, and perfect CSITR, for a SISO channel has been studied in \cite{verdu}.

\section{Optimizing energy-efficiency with no CSIT and imperfect CSIR}\label{sec:optee-nocsit}

This problem has already been well analyzed in \cite{veronica2} when perfect CSI is available at the receiver and $b=0$. So, in this paper we focus on the case when imperfect CSI is available and is obtained through channel training. For $I_{ICSIR}(P,t,\Hm)$, we use a lower bound on the mutual information obtained from the equivalent observation equation (\ref{eq:cmodel2n}), derived in \cite{mimot}:
\begin{equation}\label{eq:def-capt}
I_{ICSIR}(P,t,\hat{\Hm})=\log \left| \mb{I}_M + \frac{1}{M} \rho_{\mathrm{eff}}\left(\frac{L P}{\sigma^2} ,t\right)
 \hat{\mb{H}} \hat{\mb{H}}^H \right|
\end{equation}
Note that here, $Q=\frac{\mb{I}_M}{M}$ is used and has been shown to be optimal in \cite{veronica2}. In this section our focus is to generalize \cite{veronica2} to a more realistic scenario where the total power consumed by the transmitter (instead of the radiated power only) and imperfect channel knowledge are accounted for.

\subsection{Optimal transmit power}
\label{sec:subsec-opt-P}

By inspecting (\ref{eq:def-ee-nocsit}) and (\ref{eq:def-capt}) we see that using all the available transmit power can be suboptimal. For instance, if the available power is large and all of it is used, then $\nu_R(P,t)$ tends to zero. Since $\nu_R(P,t)$ also tends to zero when $P$ goes to zero (see \cite{veronica2}), there must be at least one maximum at which energy-efficiency is maximized, showing the importance of using the optimal fraction of the available power in certain regimes. The objective of this section is to study those aspects namely, to show that $\nu_R$ has a unique maximum for a fixed training time length and provide the equation determining the optimum value of the transmit power.

From \cite{rodriguez} we know that a sufficient condition for the function $\frac{f(x)}{x}$ to have a unique maximum is that the function $f(x)$ be sigmoidal. To apply this result in our context, one can define the function $f$ by
\begin{equation}
f(\rho_{\mathrm{eff}}) = \mathrm{Pr} \left[ \log \left| \mb{I}_M + \frac{1}{M} \rho_{\mathrm{eff}} \mb{H} \mb{H}^H \right|  \geq \xi   \right].
\end{equation}
For the SISO case, for a channel with $h$ following a complex normal distribution, it can be derived that $f(\rho)= \exp\left(- \frac{2^{\xi}-1}{\rho} \right)$ which is sigmoidal. It turns out that proving that $f$ is sigmoidal in the general case of MIMO is a non-trivial problem, as advocated by the current state of relevant literature \cite{veronica2,jorsweik,eigenrmt}. In \cite{veronica2}, $\nu_R(P)$ under perfect CSIR, was conjectured to be quasi-concave for general MIMO, and proven to be quasi-concave for the follwing special cases:
\begin{description}
  \item[(a)] $M\geq1$, $N=1$; 

  \item[(b)] $M\rightarrow +\infty$, $N < +\infty$, $\lim_{M \to \infty} \frac{N}{M}=0$;

    \item[(c)] $M < +\infty$, $N \rightarrow +\infty$, $\lim_{N \to \infty} \frac{M}{N}=0$;

  \item[(d)] $M\rightarrow +\infty$, $N \rightarrow + \infty$, $\ds{\lim_{M\rightarrow+\infty, N\rightarrow+\infty} \frac{M}{N}} = \ell < +\infty$;

  \item[(e)] $\sigma^2 \rightarrow 0$;

  \item[(f)] $\sigma^2 \rightarrow + \infty$;
\end{description}
In the following proposition, we give a sufficient condition to ensure that $\nu_R(P,t)$ is quasi-concave w.r.t $P$. 
\begin{proposition} [Optimization of $\nu_R(P,t)$ w.r.t $P$]\label{prop:qcp} If $\nu_R(P)$ with perfect CSIR is quasi-concave w.r.t $P$, then $\nu_R(P,t)$ is a quasi-concave function with respect to $P$, and has a unique maximum.
\label{tr:propforP}
\end{proposition}

This proposition is proved in Appendix \ref{sec:appproofqcp}. The above proposition makes characterizing the unique solution of $\frac{\partial  \nu_R}{\partial P}(P,t) = 0$ relevant.  This solution can be obtained through the root $\rho_{\mathrm{eff}}^*$ (which is unique because of \cite{rodriguez}) of:
\begin{equation}\label{eq:determine-P}
\frac{L}{\sigma^2}\left(P + \frac{b}{a}\right)
\frac{\tau \rho \left[(\tau+1) \rho +2 \right]}{\left[(\tau+1)^2 +1 \right]^2} f'(\rho_{\mathrm{eff}}) - f(\rho_{\mathrm{eff}}) =0
\end{equation}
with $\tau=\frac{t_s}{M}$. Note that $P$ is related to $\rho$ through $P = \sigma^2 \rho$ and $\rho$ is related to $\rho_{\mathrm{eff}}$ through (\ref{eq:rhofun}) and can be expressed as
\begin{equation}\label{eq:rho-vs-rho_eff}
 \rho  = \frac{1}{2 \tau}  \rho_{\mathrm{eff}} \sqrt{\left(1+\tau\right)^2+\frac{4 \tau}{\rho_{\mathrm{eff}}}}.
\end{equation}
Therefore (\ref{eq:determine-P}) can be expressed as a function of $\rho_{\mathrm{eff}}$ and solved numerically; once $\rho_{\mathrm{eff}}^*$ has been determined, $\rho^*$ follows by (\ref{eq:rho-vs-rho_eff}), and eventually $P^*$ follows by (\ref{eq:def-snr}). As a special case we have the scenario where $b=0$ and $\tau\rightarrow +\infty$; this case is solved by finding the unique root of $\rho^* f'(\rho^*) - f(\rho^*)=0$ which corresponds to the optimal operating SNR in terms of energy-efficiency of a channel with perfect CSI (as training time is infinite). Note that this equation is identical to that in \cite{goodman} and in this work, we provide additional insights into the form of the function $f(.)$.

Quasi-concavity is an attractive property for the energy-efficiency as quasi-concave functions can be easily optimized numerically. Additionally, this property can also be used in multi-user scenarios for optimization and for proving the existence of a Nash Equilibrium in energy-efficient power control games
\cite{goodman,lasaulce-book,lasaulce-spm}.

\subsection{Optimal fraction of training time}
\label{sec:subsec-opt-t}

The expression of $\nu_R(P,t)$ shows that only the numerator depends on the fraction of training time. Choosing $t=0$ maximizes $1-\frac{t}{T}$ but the block success rate vanishes. Choosing $t=T$ maximizes the latter but makes the former term go to zero. Again, there is an optimal trade-off to be found. Interestingly, it is possible to show that the function $\nu_R(P^*,t)$ is strictly concave w.r.t. $t$ for any MIMO channels in terms of $(M,N)$, where $P^*$ is a maximum of $\nu_R$ w.r.t $P$. This property can be useful when performing a joint optimization of $\nu_R$ with respect to both $P$ and $t$ simultaneously. This is what the following proposition states.

\begin{proposition}[Maximization of $\nu(P^*(t),t)$ w.r.t $t$]\label{prop:tcon}
The energy-efficiency function $\nu_R(P^*(t),t)$ is a strictly concave function with respect to $t$ for any $P^*(t)$ satisfying $\frac{\partial \nu_R}{\partial P}(P^*,t) =0$ and $\frac{\partial^2 \nu_R}{\partial P^2}(P^*,t) <0$, i.e, at the maximum of $\nu_R$ w.r.t. $P$.
\label{tr:propfortcc}
\end{proposition}

The proof of this proposition is provided in Appendix \ref{sec:appproofct}. The parameter space of $\nu_R$ is two dimensional and continuous as both $P$ and $t$ are continuous and thus the set $\nu(P^*(t),t)$ is also continuous and the proposition is mathematically sound. The proposition assures that the energy-efficiency can been maximized w.r.t. the transmit power and the training time jointly, provided $\nu_R(P,t)$ is quasi-concave w.r.t $P$ for all $t$. Based on this, the optimal fraction of training time is obtained by setting $\frac{\partial \nu_R}{\partial t}(P,t)$ to zero which can be written as:
\begin{equation}\label{eq:optimal-t}
 \left(\frac{T_s}{M} - \tau \right) \frac{\rho^2 (\rho+1)}{\left[\tau \rho + \rho+1 \right]^2}  f'(\rho_{\mathrm{eff}}) - f(\rho_{\mathrm{eff}}) = 0
\end{equation}
again with $\tau = \frac{t_s}{M}$. In this case, following the same reasoning as for optimizing the $\nu_R$ w.r.t. $P$, it is possible to solve numerically the equation w.r.t. $\rho_{\mathrm{eff}}$ and find the optimal $t_{s}$, which is denoted by $t^*_s$.

Note that the energy-efficiency function is shown to be concave only when it has already been optimized w.r.t $P$. The optimization problem studied here is basically, a joint-optimization problem, and we show that once $\nu(P,t)$ is maximized w.r.t $P$ for all $t$, then, $\nu(P^*(t),t)$ is concave w.r.t $t$. A solution to (\ref{eq:optimal-t}) exists only if $\nu_R$ has been optimized w.r.t $P$. However, in many practical situations, this optimization problem might not be readily solved as the optimization w.r.t $P$ for all $t$ has to be implemented first.

The following proposition describes how the optimal training time behaves as the transmit power is very large:
\begin{proposition}[Optimal $t$ in the high SNR regime]
\label{prop:optt} We have that:
$ \ds{ \lim_{P \rightarrow +\infty} t^*_s} = M $
for all MIMO systems.
\label{tr:propforthighsnr}
\end{proposition}

The proof for this can be found in Appendix \ref{sec:appproofoptt}.

\subsection{Optimal number of antennas}
\label{sec:subsec-opt-M}

So far we have always been assuming that the pre-coding matrix was chosen to be the identity matrix i.e., $\mb{Q} = \mb{I}_M$. Clearly, if nothing is known about the channel, the choice $\mb{Q} = \mb{I}_M$ is relevant (and may be shown to be optimal by formulating the problem as an inference problem). On the other hand, if some information about the channel is available (the channel statistics as far as this paper is concerned), it is possible to find a better pre-coding matrix. As conjectured in \cite{telatar} and proved in some special cases (see e.g., \cite{jorsweik}), the outage probability is minimized by choosing a diagonal pre-coding matrix and a certain number of 1's on the diagonal. The position of the 1's on the diagonal does not matter since channel matrices with i.i.d. entries are assumed. However, the optimal number of 1's depends on the operating SNR. The knowledge of the channel statistics can be used to compare the operating SNR with some thresholds and lead to this optimal number. Although we consider (\ref{eq:def-ee-nocsit}) as a performance metric instead of the outage probability, we are in a similar situation to \cite{veronica2}, meaning that the optimal pre-coding matrix in terms of energy-efficiency is conjectured to have the same form and that the number of used antennas have to be optimized. In the setting of this paper, as the channel is estimated, an additional constraint has to be taken into account that is, the number of transmit antennas used, $M$, cannot exceed the number of training symbols $t_s$. This leads us to the following conjecture.

\begin{conjecture}[Optimal number of antennas] For a given coherence time $T_s$, $\nu_R$ is maximized for $M^*=1$ in the limit of $P \to 0$. As $P$ increases, $M^*$ also increases monotonically until some $P_+$ after which, $M^*$ and $t_s^*$ decreases. Asymptotically, as $P \to \infty$, $M^*=t_s^*=1$.
\label{tr:conforM}
\end{conjecture}

This conjecture can be understood intuitively by noting that the only influence of $M$ on $\nu_R$ is through the success rate. Therefore, optimizing $M$ for any given $P$ and $t$ amounts to minimizing outage. In \cite{telatar}, it is conjectured that the covariance matrices minimizing the outage probability for MIMO channels with Gaussian fading are diagonal with either zeros or constant values on the diagonal. This has been proven for the special case of MISO in \cite{jorsweik}, we can conclude that the optimal number of antennas is one in the very low SNR regime and that it increments as the SNR increases. However, the effective SNR decreases by increasing $M$ (seen from expression of $\rho_{\mathrm{eff}}$ and $\tau$) , this will result in the optimal $M$ for each $P$ with training time lower than or equal to the optimal $M$ obtained with perfect CSI. Concerning special cases, it can be easily verified that the optimal number of antennas is $1$ at very low and high SNR.

At last, we would like to mention a possible refinement of the definition in (\ref{eq:def-ee-nocsit}) regarding $M$. Indeed, by creating a dependency of the parameter $b$ towards $M$ one can better model the energy consumption of a wireless device. For instance, if the transmitter architecture is such that one radio-frequency transmitter is used per antenna, then, each antenna will contribute to a separate fixed cost. In such a situation the total power can written as $aP+Mb_0$ where $b_0$ is the fixed energy consumption per antenna. It can be trivially seen that this does not affect the goodput in any manner and only brings in a constant change to the total power as long as $M$ is kept a constant. Therefore, the optimization w.r.t $P$ and $t$ will not change but it will cause a significant impact on the optimal number of antennas to use.

\section{Numerical results and interpretations}\label{sec:numerical-study}

We present several simulations that support our conjectures as well as expand on our analytical results. All simulations are performed using Monte-Carlo simulations as there is no expression available for the outage of a general MIMO system.

\subsection{With imperfect CSITR available}

The $F_L$ we use here is based on the results in \cite{Buckingham-08}, $F_L= Q_{func}(\frac{\xi-I_{\mathrm{ICSITR}}(P,\Qm_{WF},\Hm)}{\sqrt{\frac{2\rho}{(1+\rho)L} } })$, $L$ being the code-length. This is the Gaussian approximation that is very accurate for $L$ large enough and from simulations we observe that for $L \geq 10$ the approximation is quite valid.

\begin{figure}[h]
    \begin{center}
        \includegraphics[width=180mm]{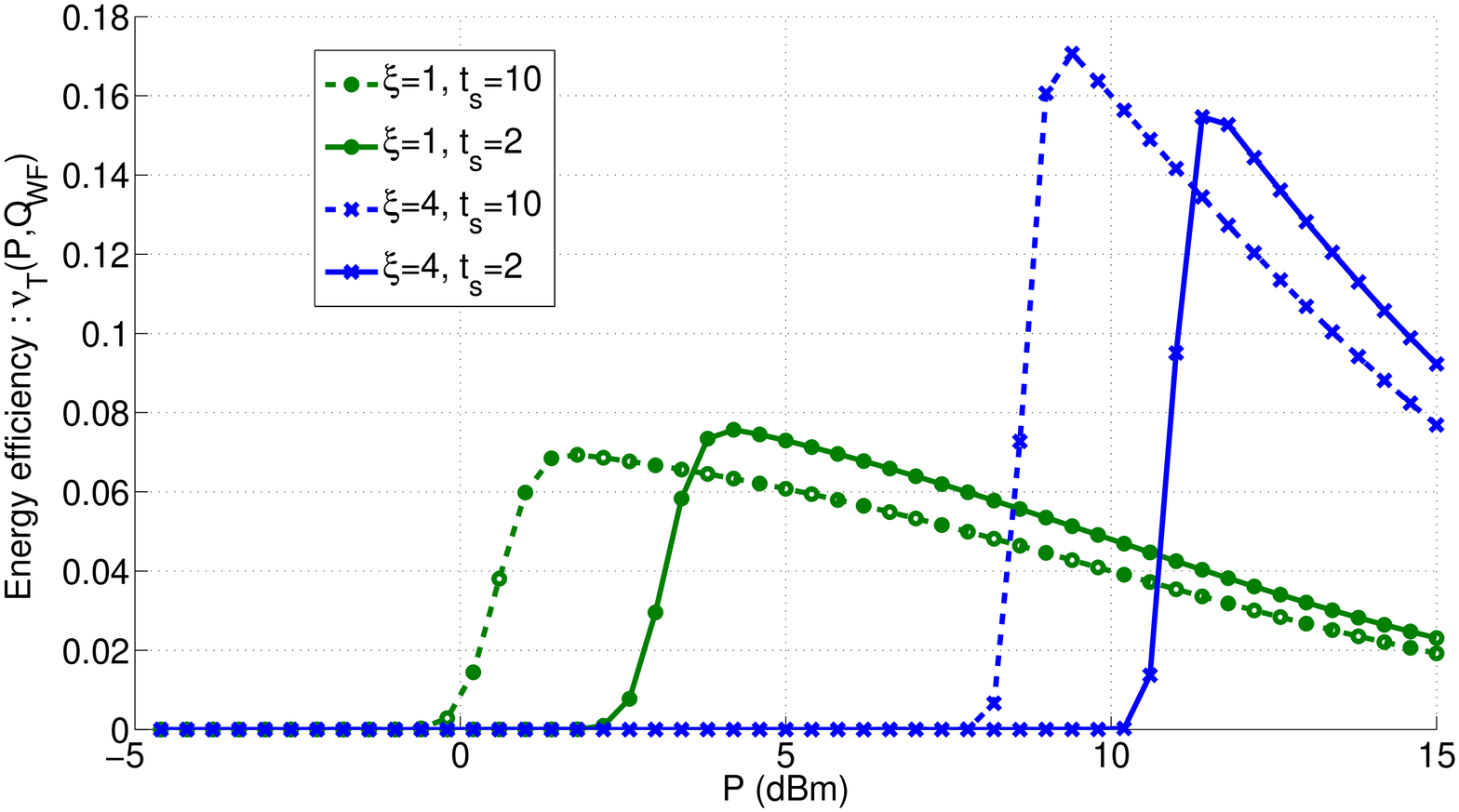}
    \end{center}
    \caption{Energy efficiency ($\nu_T$) in bits/J v.s transmit power (P) in dBm for a MIMO system with imperfect CSITR, $M=N=2$, $R_0=1$bps, $T_s=100$, $\frac{b}{a}=10$ mW for certain values of $\xi$ and $t_s$.}
    \label{fig:eff_rho_csit}
\end{figure}

\begin{figure}[h]
    \begin{center}
        \includegraphics[width=180mm]{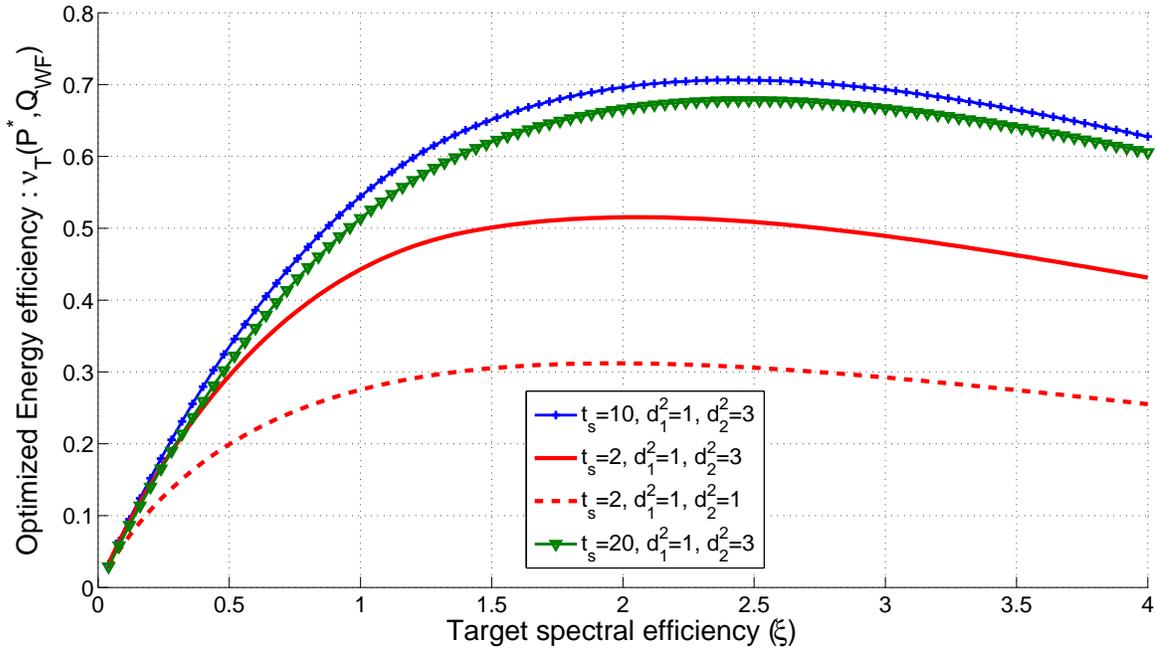}
    \end{center}
 \caption{Optimal energy-efficiency ($\nu_T(P^*,\Qm_{WF})$) in bits/J v.s spectral efficiency ($\xi$) for a MIMO system with imperfect CSITR, $M=N=2$, $R_0=1$bps, $T_s=100$, $L=100$ and $\frac{b}{a}=1$ mW.}
    \label{fig:eff_xi_b1}
\end{figure}

\begin{figure}[h]
    \begin{center}
        \includegraphics[width=180mm]{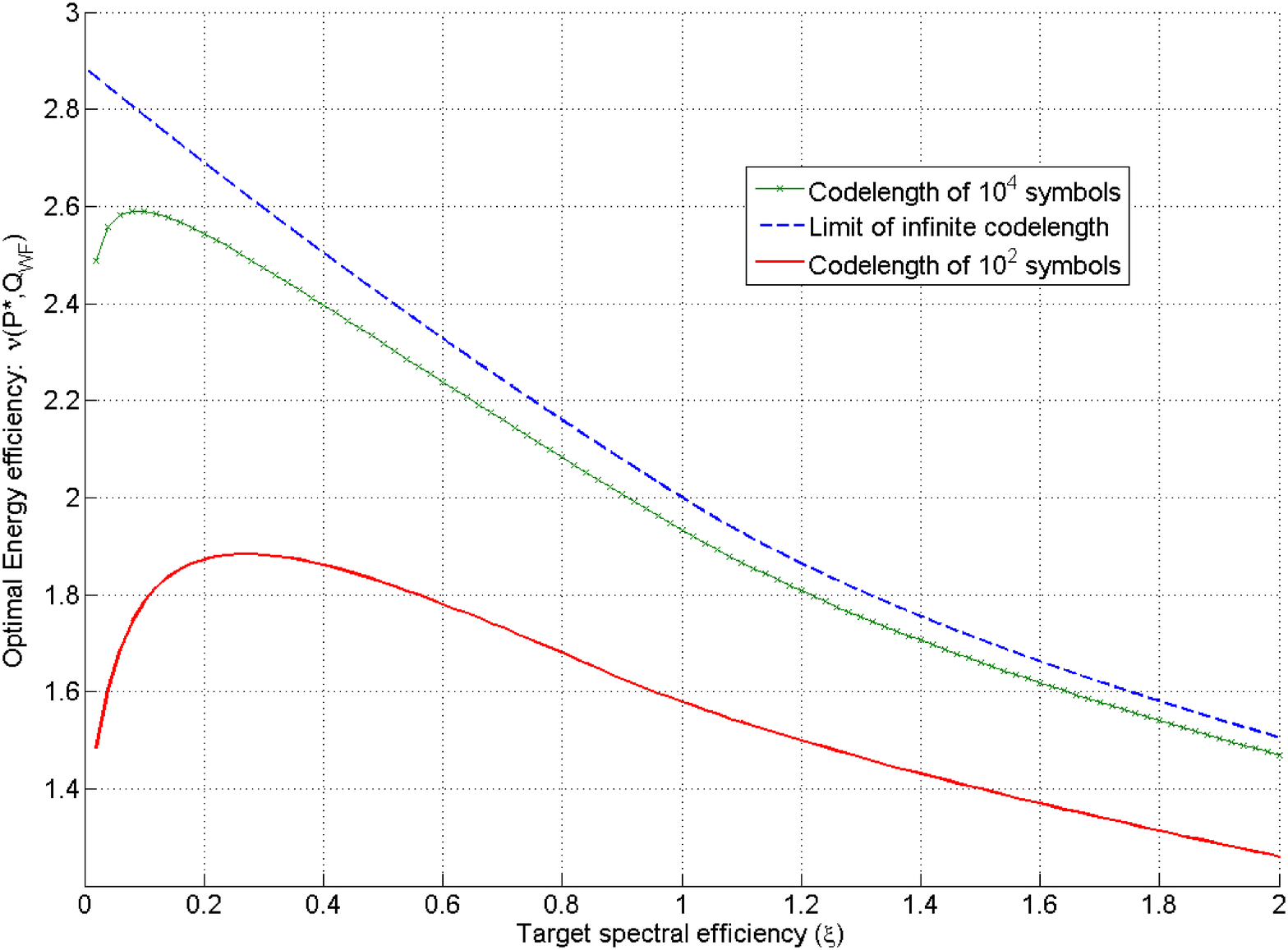}
    \end{center}
    \caption{Optimal energy-efficiency ($\nu_T(P^*,\Qm_{WF})$) v.s spectral efficiency ($\xi$) for a MIMO system with perfect CSITR, $M=N=2$, $R_0=1$bps and $\frac{b}{a}=0$.}
    \label{fig:eff_xi_b0}
\end{figure}

First of all, numerical results are presented that support and present our analytical results through figures. The first two figures shown assume imperfect CSITR obtained through training and use a $2 \times 2$ MIMO system. The quasi-concavity of the the energy-efficiency function w.r.t the transmit power is shown in Figure \ref{fig:eff_rho_csit} for $\xi=1$ and $\xi=4$, and $t_s=2$ and $t_s=10$. This figure shows that for a higher target rate, a longer training time yields a better energy-efficiency. We also observe that using a higher $\xi$ can results in a better energy-efficiency as in this figure. This motivates us to numerically investigate if there is also an optimal spectral efficiency to use, given a certain $T_s$, $\frac{b}{a}$ and $L$. Figures \ref{fig:eff_xi_b0} and \ref{fig:eff_xi_b1} present the results of this study.

Surprisingly, we observe that our plots are quasi-concave and so there is an optimal target rate to use for each channel condition and code-length. In Figure \ref{fig:eff_xi_b1}, $\nu_T$ is always optimized over $P$ and $\Qm$. Observe that $\nu_T^*(\xi)$ is also quasi-concave and has a unique maximum for each value of $d_i$ and $t_s$ (representing the channel Eigen-values as from equations (\ref{eq:def-cappcsitmod}), (\ref{eq:wf})  and training time lengths). $d_i$ is ordered in an ascending order, i.e. in this case, with $d_1^2 \leq d_2^2$. The parameters used are: $M=N=2$, $R_0=1$bps, $T_s=100$, $L=100$ and $\frac{b}{a}=1$ mW with $t_s=2,10$ and $20$ for $d_1^2=1$, $d_2^2=3$, and $t_s=2$ for $d_1^2=d_2^2=1$.This figure also implies that the training time and target rate can be optimized to yield the maximum energy-efficiency for a given coherence-time and channel fading. For infinite code-length the plot is maximized at the solution of (\ref{eq:ee-infcl}). While for Figure \ref{fig:eff_xi_b0}, perfect CSIT is assumed with $b=0$, at infinite block length, the optimal transmit rate/power is zero as expected (also seen from (\ref{eq:ee-infcl})). However, remarkably, for finite code-lengths there is a non-zero optimal rate and corresponding optimal power as seen from the figure.

\begin{figure}[h]
    \begin{center}
        \includegraphics[width=180mm]{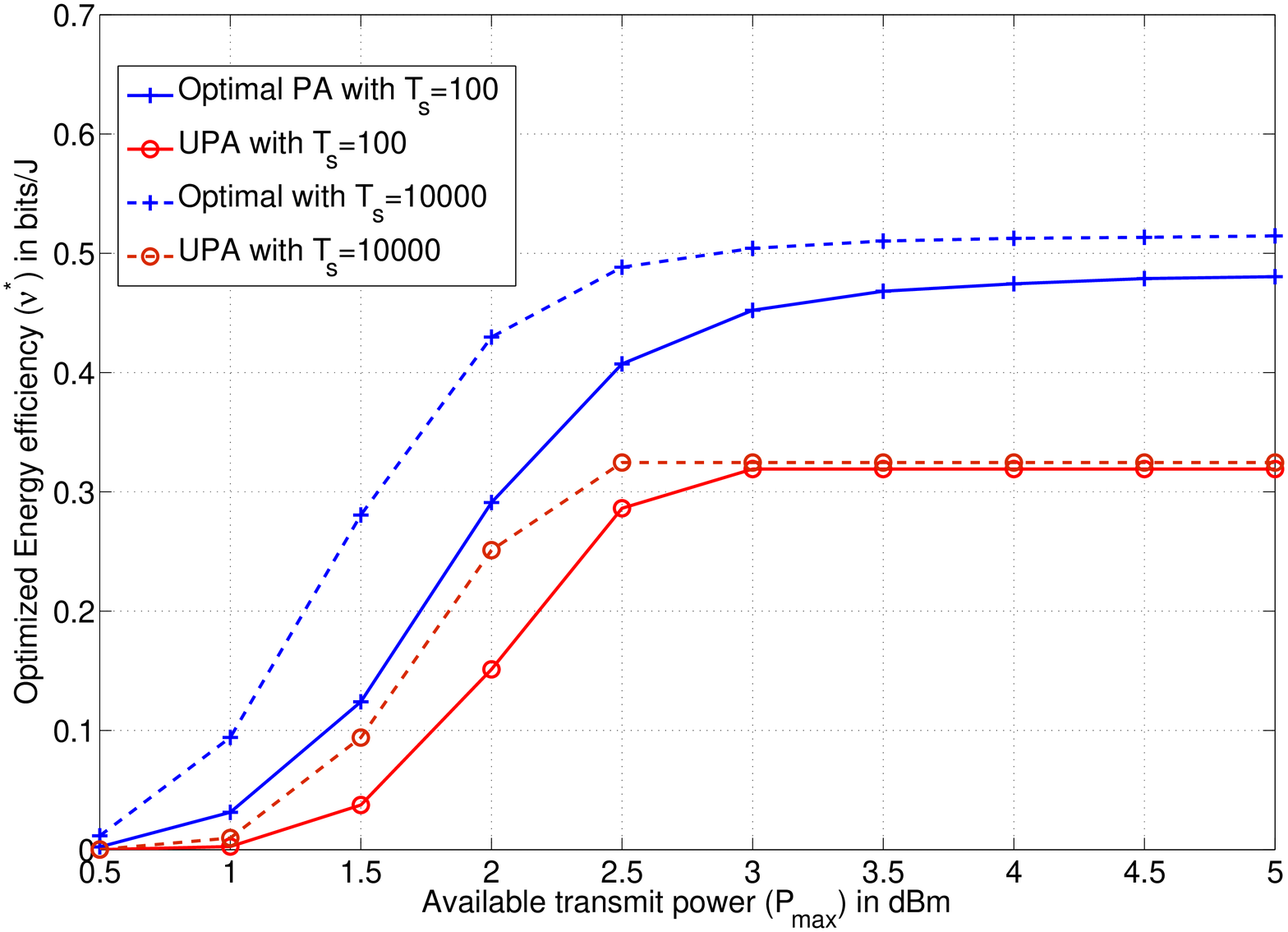}
    \end{center}
    \caption{Optimal energy-efficiency ($\nu_R(P^*)$) in bits/J v.s available transmit power ($\sup(P)$) for a MIMO system with imperfect CSITR, $M=N=2$, $R_0=1$bps, $R=1$bps and $\frac{b}{a}=1$ mW.}
    \label{fig:eff_comp_pa}
\end{figure}

Finally in Figure \ref{fig:eff_comp_pa}, we compare our energy efficiency function that uses optimized power allocation to uniform power allocation, and present the gain from having CSIT. In both cases, the training time and the transmit power is optimized and we plot the optimized energy efficiency v.s $P_{\max}$. Note that the optimized PA always yields a better performance when compared to UPA and at low power, UPA has almost zero efficiency while the optimal PA yields a finite efficiency. The gain observed can be considered as the major justification in using non-uniform power allocation and sending the channel state information to the transmitter. However, when the block length is small, imperfect CSIT results in a smaller gain as seen from the relatively larger gap between $T_s=100$ and $T_s=10000$ when compared to the size of the gap in UPA.


\subsection{With no CSIT}

We start off by confirming our conjecture that for a general MIMO system, $\nu_R(P,t)$ has a unique maximum w.r.t $P$. We also confirm that optimal values of training lengths and transmit antennas represented by $t^*_s$ and $M^*$ are as conjectured.\\

Once the analytical results have been established, we explore further and find out the optimal number of antennas and training time when $\nu_R$ has been optimized w.r.t $P$. For this we use the optimized energy efficiency defined as $\nu^*(P,t)=\max\{\nu_R(p,t) \| p \in [0,P] \}$. As we know $\nu_R$ to be quasi-concave w.r.t $P$ and having a unique maximum, this newly defined $\nu^*$ will indicate what is the best energy efficiency achievable given a certain amount of transmit power $P$. Hence, plotting $\nu^*$ against $P$ for various values of $M$ or $t_{s}$ can be useful to determine the optimal number of antennas and training time while using the optimal power.

In the following plots we take $\frac{\sigma^2}{L}=1mW$ so that $P$ can be expressed in $dBm$ easily. Also note that $\frac{b}{a}$ has the unit of power and is expressed in Watts (W). We also use $S_d = 15$ $\mu$s from LTE standards \cite{lte}.

\begin{figure}[h]
    \begin{center}
        \includegraphics[width=180mm]{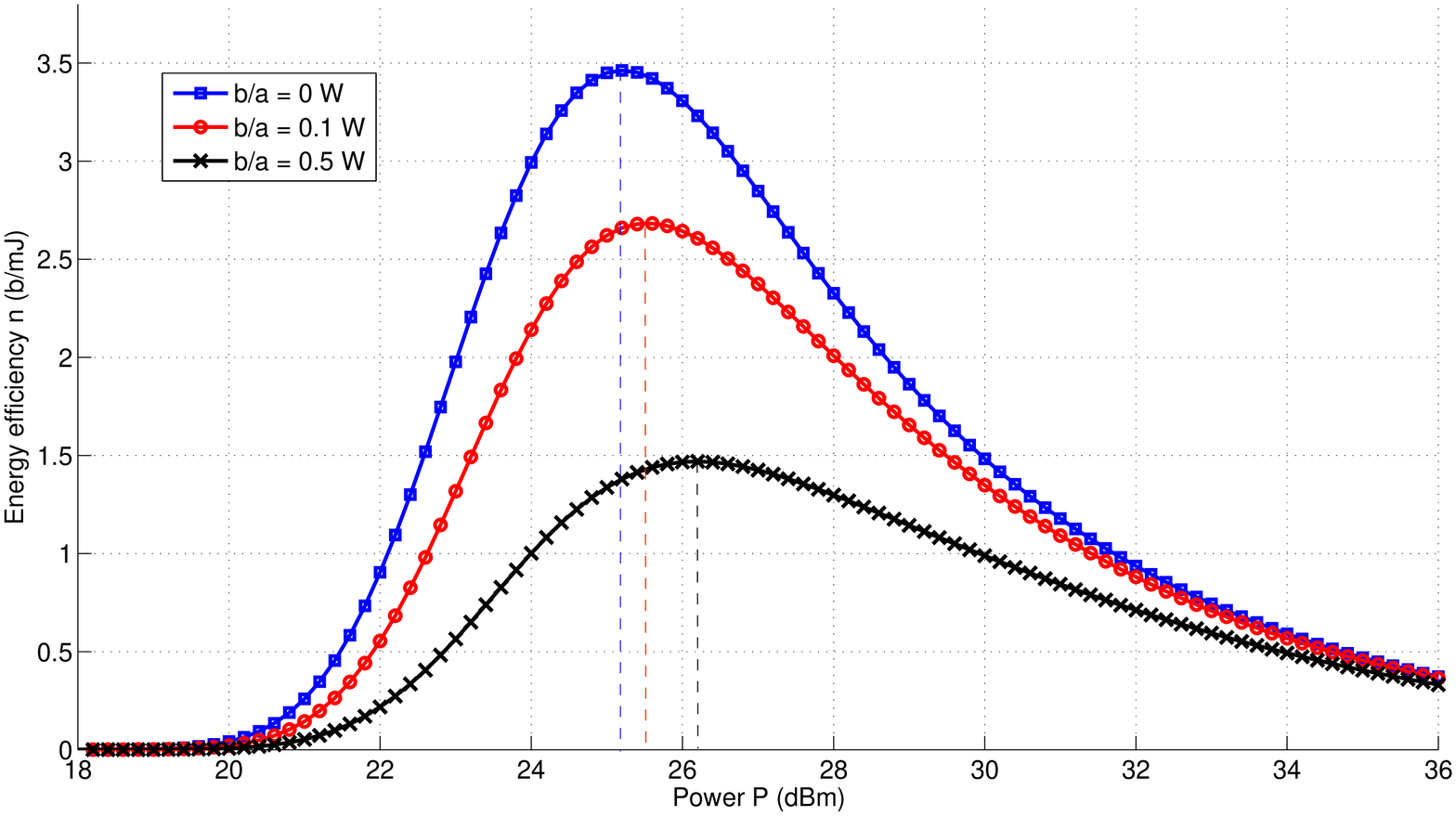}
    \end{center}
    \caption{Energy efficiency ($\nu_R$) v.s transmit power (P) with $t_{s}=M=N=4$,$R=1600$bps, $\xi=\frac{R}{R_0}$=16 and $T_s=55$ symbols.}
    \label{fig:eff_rho}
\end{figure}

Figure \ref{fig:eff_rho} studies the energy efficiency as a function of the transmit power ($P$) for different values of $\frac{b}{a}$ and illustrates the quasi-concavity of the energy efficiency function w.r.t $P$. The parameters used are $R=1600$, $\xi=\frac{R}{R_0}=16$, $T_s=55$ and $M=N=t=4$.

\begin{figure}[h]
    \begin{center}
        \includegraphics[width=180mm]{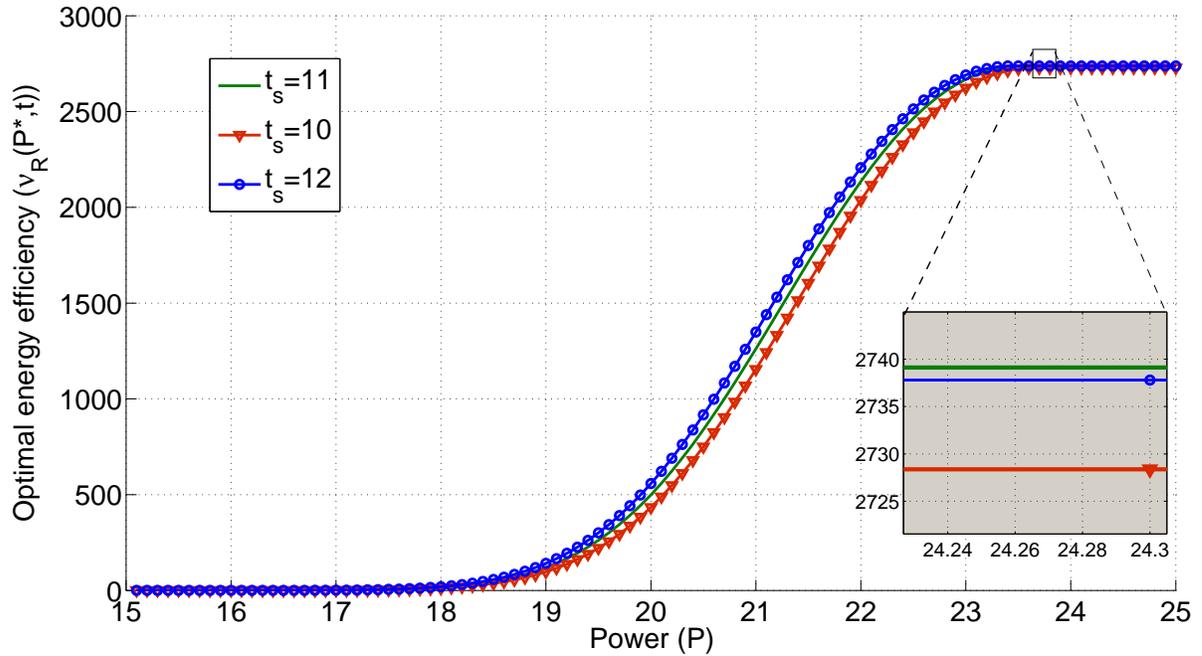}
    \end{center}
    \caption{Optimized efficiency ($\nu^*$) vs. maximum transmit power ($P$) for a MIMO system with $M=N=4$, $R = 1$ Mbps, $\xi=\frac{R}{R_0}$=16, $T_s=55$ and $\frac{b}{a}=0$W.}
    \label{fig:effz}
\end{figure}

Figure \ref{fig:effz} studies the optimized energy efficiency $\nu^*$ as a function of the transmit power with various values of $t_{s}$. The figure illustrates that beyond a certain threshold on the available transmit power, there is an optimal training sequence length that has to be used to maximize the efficiency, when the optimization w.r.t $P$ has been done, which has been proven analytically in proposition \ref{prop:tcon}.  The parameters are $R=1$Mbps, $\xi=16$, $\frac{b}{a}=0$, $M=N=4$, $\frac{b}{a}=0$ and $T_s=55$.

\begin{figure}[h]
    \begin{center}
      \includegraphics[width=180mm]{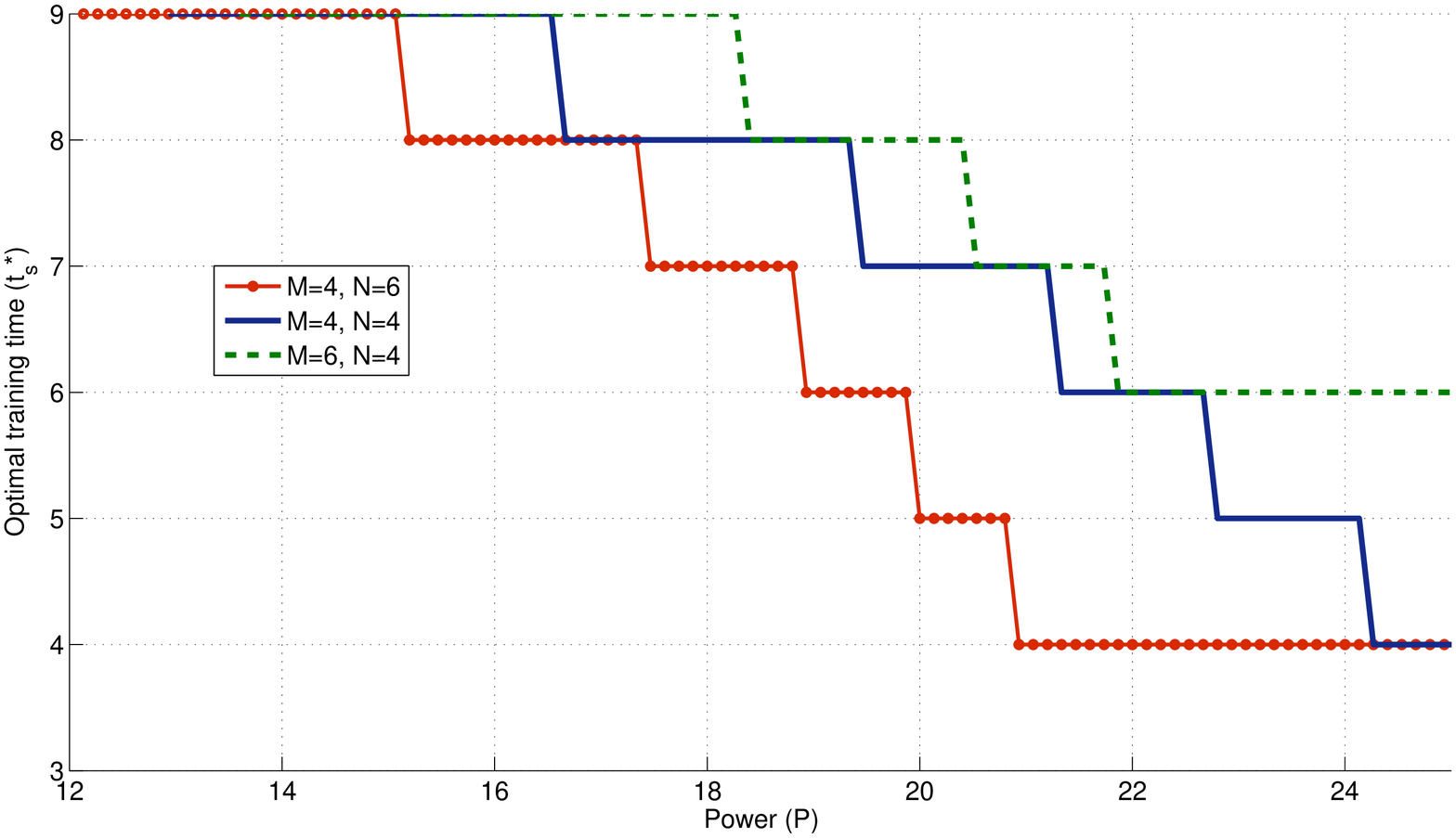}
   \end{center}
    \caption{Optimal training sequence length ($t_{s}$) vs. Transmit Power (P) MIMO system with $\xi=\frac{R}{R_0}=16$,$R=1$Mbps, $N=4$, $T_s=10$ symbols. The discontinuity is due to the discreteness of $t_s$.}
   \label{fig:eff_t}
\end{figure}

Figure \ref{fig:eff_t} studies the optimal training sequence length $t_{s}$ as a function of the transmit power $P$. Note that in this case, we are not optimizing the efficiency with respect to $P$ and so this figure illustrates proposition \ref{tr:propforthighsnr}. With $P$ large enough $t_{s}=M$ becomes the optimal training time and for $P$ small enough $t_{s}=T_s-1$ as seen from the figure. The parameters are $R=1600$, $\frac{b}{a}=0$ W, $\xi=16$ and $T_s=10$. (We use $T_s=10$, as if the coherence time is too large, the outage probabilities for low powers that maximize the training time, such that $t^*_s=T_s-1$, become too small for any realistic computation.)

\begin{figure}[h]
    \begin{center}
        \includegraphics[width=180mm]{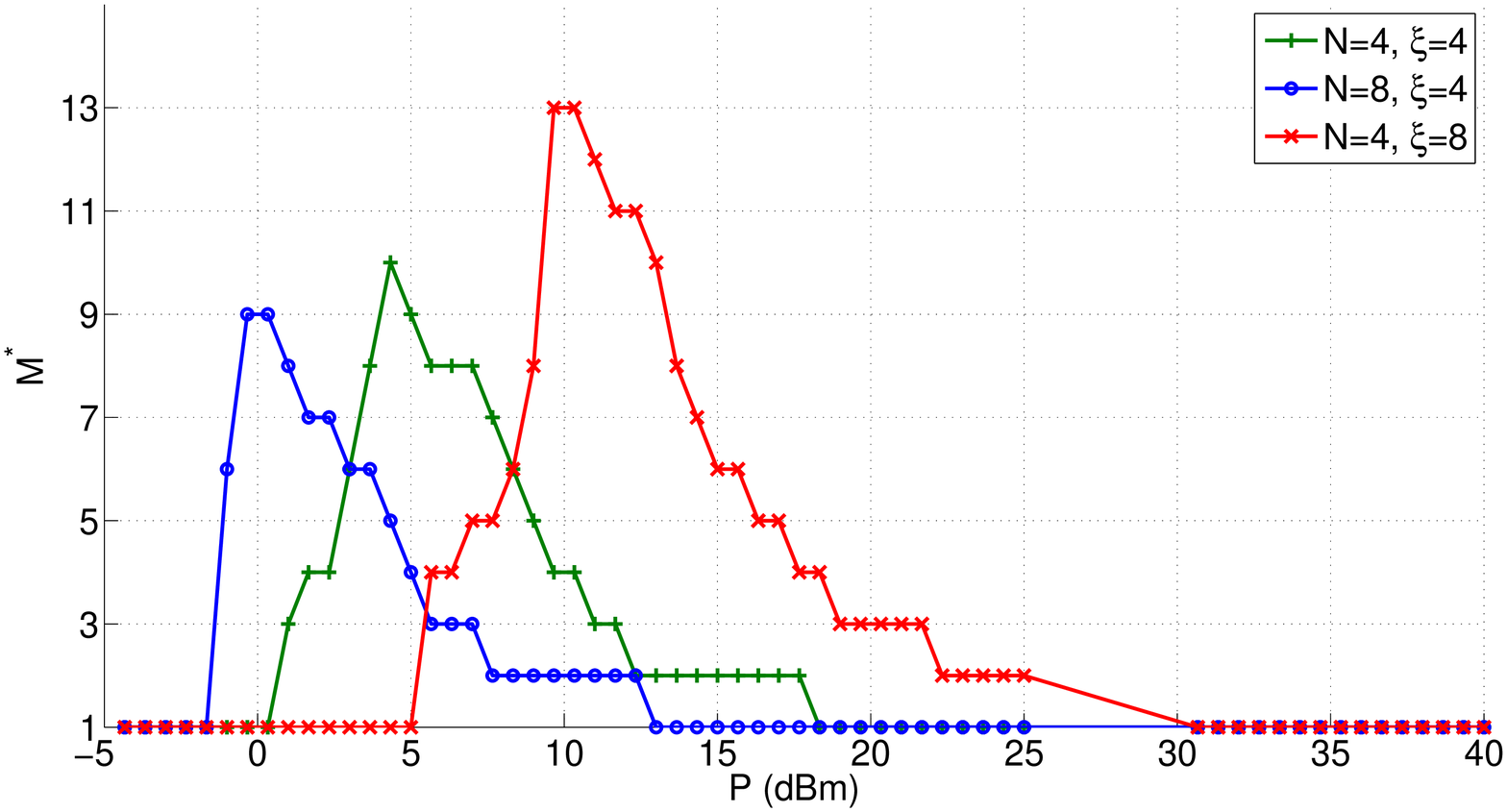}
    \end{center}
    \caption{Optimal number of antennas ($M^*$) vs. Transmit Power (P) in dBm for a MIMO system with $R_0=1$Mbps, $T_s=100$, for certain values of $\xi$ and $N$, and $t_s$ optimized jointly with $M$. The discontinuity is due to the discreteness of $M$.}
    \label{fig:eff_fixt}
\end{figure}

Figure \ref{fig:eff_fixt} studies the optimal number of antennas $M^*$ as a function of the transmit power $P$ with the training time optimized jointly with $M$. With $P$ large enough $M=t_{s}=1$ becomes the optimal number of antennas and for $P$ small enough $M=1$ as seen from the figure. This figure illustrates conjecture \ref{tr:conforM}. The parameters are $R_0=1$Mbps, $\frac{b}{a}=10$ mW and $T_s=100$.

From all of our theoretical and numerical results so far, we can conclude that given a target spectral efficiency $\xi$, a coherence block length $T_s$ and number of receive antennas, there is an optimal transmit power $P^*$, transmit antennas $M^*$ and training time $t^*_s$ to use that optimizes the energy efficiency.


\section{Conclusion}\label{sec:conclusion}

This paper proposes a framework for studying the problem of energy-efficient pre-coding (which includes the problem of power allocation and control) over MIMO channels under imperfect channel state information and the regime of finite block length. As in \cite{goodman}, energy-efficiency is defined as the ratio of the block success rate to the transmit power. But, in contrast with \cite{goodman} and the vast majority of works originating from it, we do not assume an empirical choice for the success rate such as taking $f(x) = (1 - e^{-x})^L$, $L$ is the block length. Instead, the numerator of the proposed performance metric is built from the notion of information, and more precisely from the average information (resp. mutual information) in the case where CSIT is available (resp. not available). This choice, in addition to giving a more fundamental interpretation to the metric introduced in \cite{goodman}, allows one to take into account in a relatively simple manner effects of practical interest such as channel estimation error and block length finiteness. Both in the case where (imperfect) CSIT is available and not available, it is shown that using all the available transmit power is not optimal. When CSIT is available, whereas determining the optimal power allocation scheme is a well known result (water-filling), finding the optimal total amount of power to be effectively used is a non-trivial choice. Interestingly, the corresponding optimization problem can be shown to be quasi-convex and have a unique solution, the latter being characterized by an equation which is easy to solved numerically. When CSIT is not available, solving the pre-coding problem in the general case amounts to solving the Telatar's conjecture. Therefore, a new conjecture is proposed and shown to become a theorem in several special cases. Interestingly, in this scenario, it is possible to provide a simple equation characterizing the optimal fraction of training time. Numerical results are provided to sustain the proposed analytical framework, from which interesting observations can be made which includes~: block length finiteness gives birth to the existence of a non-trivial trade-off between spectral efficiency and energy efficiency~; using optimal power allocation brings a large gain in terms of energy-efficiency only when the channel has a large enough coherence time,demonstrating the value of CSIT and channel training.

The proposed framework is useful for engineers since it provides considerable insights into designing the physical layer of MIMO systems under several assumptions on CSI. The proposed framework also opens some interesting research problems related to MIMO transmission, which includes~: finding the optimal pre-coding matrix for the general case of i.i.d. channel matrices under no CSIT. Even in the case of large MIMO systems, this problem is not solved~; extending the proposed approach to the case of Rician channels with spatial correlations~; tackling the important case of multiuser MIMO channels~; considering the problem of distributed energy-efficient pre-coding.


\appendices

\section{Proof of theorem \ref{th:qcwf}}
\label{sec:proofqcwf}

In order to prove that $\nu_T(P,\Qm_{WF(P)},\Hm) $ is quasi-concave with respect to $P$ and has a unique maximum $\nu_T(P^*,\Qm_{WF(P^*)},\Hm)$, we exploit the result in \cite{rodriguez} which states that if $f(x)$ is an ``S"-shaped or sigmoidal function, then $\frac{f(x)}{x}$ is a quasi-concave function with a unique maximum. An ``S"-shaped or sigmoidal function has been defined in \cite{rodriguez} in the following manner. A function $f$ is ``S" shaped, if it satisfies the following properties:
\begin{enumerate}
\item Its domain is the interval $[0, \infty)$.
\item  Its range is the interval $[0,1)$.
\item It is increasing.
\item (``Initial convexity") It is strictly convex over the interval $[0, x_f ]$, with $x_f$ a positive number.
\item (``Eventual concavity") It is strictly concave over any interval of the form $[x_f, L]$, where $x_f<L$.
\item It has a continuous derivative.
\end{enumerate}

Considering the non-constant terms in $\nu_T$, we see that what we have to show is that $F_L(I_{\mathrm{ICSITR}}(P,\Qm_{WF(P)},\Hm)-\xi)$ is ``S"-shaped w.r.t $P$. We already have that $F_L(x)$ is sigmoidal, therefore all we have to show is that $F_L(g(P))$ is also sigmoidal where $g(P)= I_{\mathrm{ICSITR}}(P,\Qm_{WF(P)},\Hm)-\xi$. Trivially, when $P=0$, $F_L(I_{\mathrm{ICSITR}}(P))=0$ and $\lim_{P \to \infty} F_L=1$. The rest can be proved using the following arguments:
\begin{itemize} 
\item $g(P)$ is continuous: As $P$ varies, $\Qm_{WF(P)}$ also is modified according to the iterative water-filling algorithm. This results in using one antenna for low $\rho$ to all antennas for high values of $\rho$.

There exists certain ``threshold" points of the total power, $P^{\mathrm{th}}_{i}, i=\{ 1,\dots,M \} $, at which the number of antennas used changes. The convention being, for $P^{\mathrm{th}}_{i-1} \leq P \leq P^{\mathrm{th}}_{i}$, $i$ number of antennas are used ($s$ for the rest are set to zero). $P^{\mathrm{th}}_{0}=0$ and $P^{\mathrm{th}}_{M}=\infty$ . If $I_{\mathrm{CSITR}}(P,\Qm_{WF(P)},\Hm)$ is continuous at these points, then $g(P)$ is continuous. It can also be observed that in all other points, $I_{\mathrm{CSITR}}(P,\Qm_{WF(P)},\Hm)$ can be expressed as $\Sigma_{i=1} ^{J} \log(1+\alpha_i+\beta_i s_i), J \leq \min(M,N)$. ($\alpha$ and $\beta$ is obtained from solving (\ref{eq:wf}).)
A ``threshold" point occurs when $P=P_j^{\mathrm{th}}$ $s_j=0$ is obtained by solving (\ref{eq:wf}). The left hand limit is that $j-1$ antennas are used and so, $I_{\mathrm{CSITR}}(P,\Qm_{WF(P)},\Hm)=\Sigma_{i=1}^{j-1}\log\left(1+\alpha_i+\beta_i s_i\right)$. The right hand limit will be obtained by solving (\ref{eq:wf}), with $s_1 \to 0$ (assuming without loss of generality that $d_1^2$ is the smallest). This will yield a solution which can be easily seen to be the same as the left hand limit as $p_1 \to 0$.

\item We have shown $g(P)$ to be a finite sum of logarithms of a monomial expansion of $P$ in certain intervals (marked by $P_i^{\mathrm{th}}$) . For each interval it is trivial to see that $F_L(g(P))$ is also ``S"-shaped. As $g(P)$ is continuous, $F_L(g(P))$ is ``S"-shaped for all $P$. 
\item From Lemma \ref{aux:sshaped} proved in Appendix \ref{sec:appproofqcp} we can show that $\frac{F_L(g(P))}{aP+b}$ is also ``S"-shaped by a simple change of variable $x=aP+b$. Thus, we have $\nu_T(P,\Qm_{WF(P)},\Hm)$ as a quasi-concave function with a unique maximum.

\item With imperfect CSI, the only change is in $I_{\mathrm{ICSITR}}(P^*,\Qm_{WF(P^*)},\widehat{\Hm})$ now given from (\ref{eq:def-capipcsit}). The water-filling algorithm now replaces $\Hm$ with $\widehat{\Hm}$ and so on. This maintains the continuity of $g(P)$. However we now have $I_{\mathrm{ICSITR}}(P,\Qm_{WF(P)},\widehat{\Hm})=  \Sigma_{i=1}^{j}\log\left(1+\frac{\alpha_i+\beta_i s_i}{1+\rho \sigma_E^2}\right) $. From \cite{mimot} we have $\frac{1}{1+\rho \sigma_E^2}$ as a concave function and so even in this case, we have $F_L(g(P))$ as a sigmoidal function and  $\nu_{T}(P,\Qm_{WF(P)},\widehat{\Hm}) $ as quasi-concave with a unique maximum.
As it is continuous and differentiable, the maximum can be found as the unique solution to the equation:
\begin{eqnarray}
&\frac{ \partial F_L[I_{\mathrm{ICSITR}}(P^*,\Qm_{WF(P^*)},\widehat{\Hm})-\xi]}{\partial P}\left(P^*+\frac{b}{a}\right) & \\ \nonumber  
&-F_L[I_{\mathrm{ICSITR}}(P^*,\Qm_{WF(P^*)},\Hm)-\xi] & =0
\end{eqnarray}
where $\frac{\partial}{\partial P} $ is the partial derivative.
\end{itemize}

QED
\section{Proof of proposition \ref{prop:qcp}}
\label{sec:appproofqcp}
An ``S"-shaped function has been defined in \cite{rodriguez} in the following manner. A function $f$ is ``S" shaped, if it satisfies the properties as mentioned in Appendix \ref{sec:proofqcwf}.
\begin{Lemma}
\label{aux:sshaped}
If $f$ is a ``S" shaped function, the composite function $f\circ g (x)$ is also ``S" shaped if $g$ satisfies the following properties:
\begin{enumerate}
\item $g$ also satisfies conditions 1, 3, 4 and 6 but with $g(0)=b, b>0$.
\item $\ds{\lim_{x \to \infty}} f'(x)g'(x)=0$.
\item $g''(x)$ is a decreasing function such that $\ds{\lim_{x \to \infty}} g''(x) = 0$.

\end{enumerate}
\end{Lemma}
The proof for the above Lemma is at the end of this section.

In \cite{veronica2}, the authors prove that the energy efficiency function with perfect CSI defined as the goodput ration to transmitted RF signal power is a quasi concave function by showing that the success rate function, $f(\rho)$ is ``S" shaped for the following cases:
\begin{description}
  \item[(a)] $M\geq1$, $N=1$;

  \item[(b)] $M\rightarrow +\infty$, $N < +\infty$;

    \item[(c)] $M < +\infty$, $N \rightarrow +\infty$;

  \item[(d)] $M\rightarrow +\infty$, $N \rightarrow + \infty$, $\ds{\lim_{M\rightarrow+\infty, N\rightarrow+\infty} \frac{M}{N}} = \ell < +\infty$;

  \item[(e)] $\sigma^2 \rightarrow 0$;

  \item[(f)] $\sigma^2 \rightarrow + \infty$;
\end{description}
So, if we can show that the success rate function in our situation is also ``S" shaped, our proof is complete for all the cases mentioned above. From (\ref{eq:cmodel2n}) we know that the worst case mutual information in the case of imperfect CSI with training is mathematically equivalent to that of perfect CSI but with $\rho$ replaced by $\rho_{\mathrm{eff}}$. Thus it is possible to replace $f(\rho)$, in the case of perfect CSI, by $f(\rho_{\mathrm{eff}})$, when we study the case of imperfect CSI, and so we can study the energy efficiency function given by:
\begin{eqnarray}
\nu_R(P,t) &=& R\zeta\frac{f(\rho_{\mathrm{eff}}(p(x)))} {x}
          \label{eq:fofnu}
\end{eqnarray}
where $x$ is a new variable that represents the total consumed power and $p(x) = \frac{L(x-b)}{a\sigma^2}$. $p'(x)>0$ and $p''(x)=0$ and $\rho_{\mathrm{eff}}'(\rho)>0$ and $\ds{\lim_{\rho \to \infty}} \rho_{\mathrm{eff}}''(\rho)=0$. Thus $\rho_{\mathrm{eff}}$ and $p$ satisfy the conditions on $g$ detailed in Lemma \ref{aux:sshaped}. Hence we have proven that the numerator is ``S" shaped with respect to $x$ and then it immediately follows from the results in \cite{rodriguez} that $\nu_R$ has a unique maximum and is quasi-concave for all the specified cases.

\bf Proof of Lemma \label{aux:sshaped} \normalfont

Here we show that $f \circ g$ also satisfies all the properties of the ``S" function as described in \cite{rodriguez}.
\begin{enumerate}
\item Its domain is the domain of $g$ which is clearly the non-negative part of the real line; that is, the interval $[0,\infty)$.
\item  Its range is the range of $f$, the interval $[0,1)$.
\item  It is increasing as both $f$ and $g$ are increasing.
\item (``Initial convexity"). Note that $f(g(x))''=f''(y)g'(x)+g''(x)f'(y)$, with $y=g(x)$. As all terms in this expansion are positive in the interval $[0, x_f ]$, $f\circ g$ is also convex in this interval. Also note that as $g'$ and $f'$ are strictly positive and $g''$ is decreasing, thus for $y>x_f$ once $f(g(x))''<0$ it stays negative till infinity. This implies that if there is an inflexion point, it is unique.
\item (``Eventual concavity") Consider $h(x)=f(g(x))'=f'(y)g'(x)$, due to the initial convexity and increasing nature of $h$, $h(x_f)=k, k>0$. $\ds{\lim_{x \to \infty}} f(g(x_f))'=0$. As $h$ is continuous the mean value theorem imposes $h'(x)<0$ at some point. This implies that there exists some point $x_d > 0$ such that $f\circ g$ is concave in the interval $[x_d, \infty$] and convex before it.
\item It has a continuous derivative. (all the functions used here are continuous)
\end{enumerate}
Hence, $f\circ g$ is ``S" shaped.

QED

\section{Proof of proposition \ref{prop:tcon}}
\label{sec:appproofct}
Let us consider the second partial derivative of $\nu_R$ with respect to $t$. (Note that this is possible as $t$ is a real number with the unit of time while $t_{s}$ is a natural number.) From (\ref{eq:fofnu}), $\nu_R(P,t)=K^{-1} (1-\frac{t}{T})f(\rho_{\mathrm{eff}})$, with $K^-1=\frac{R}{x}$ a constant if $P$ is held a constant.

\begin{eqnarray}
\label{eq:nutty}
\frac{K\partial^2 \nu}{\partial t^2} = & (1-\frac{t}{T})f''(\rho_{\mathrm{eff}})\rho_{\mathrm{eff}}'(t)^2 +(1-\frac{t}{T})f'(\rho_{\mathrm{eff}})\rho_{\mathrm{eff}}''(t) & \nonumber\\ 
&- \frac{2}{T}f'(\rho_{\mathrm{eff}})\rho_{\mathrm{eff}}'(t) &
\end{eqnarray}

In the above sum, it can be easily verified that the terms $f'(\rho_{\mathrm{eff}})\rho_{\mathrm{eff}} '(t)$ and $f'(\rho_{\mathrm{eff}})$ are positive and that $\rho_{\mathrm{eff}} ''(t)<0$. Thus if we have $f''(\rho_{\mathrm{eff}}) <0$, then $\nu_R(t)$ is strictly concave.

The only way $\nu_R$ depends on $P$ is through $\frac{f(\rho_{\mathrm{eff}})}{aP+b}$ and $R\zeta$ stays a constant if only $P$ changes. So if we use the fact that we are working at a maximum of $\nu_R$ with respect to $\rho$, i.e $\frac{\partial \nu}{\partial \rho}=0$ and $\frac{\partial^2 \nu}{\partial \rho ^2}<0$, we have $\frac{\partial \nu}{\partial \rho}$ as:

\begin{equation}
\label{eq:nudrho1}
0= f'(\rho_{\mathrm{eff}})\rho_{\mathrm{eff}}'(\rho)\rho^{-1} - f(\rho_{\mathrm{eff}})\rho^{-2}
\end{equation}

And, substituting (\ref{eq:nudrho1}) in $\frac{\partial^2 \nu}{\partial \rho^2}<0$, that is:

\begin{eqnarray}
f''(\rho_{\mathrm{eff}})(\rho_{\mathrm{eff}}')^2\rho^{-1}-2f'(\rho_{\mathrm{eff}})\rho_{\mathrm{eff}}'\rho^{-2} + 2f(\rho_{\mathrm{eff}})\rho^{-3} &<0& \nonumber \\
f''(\rho_{\mathrm{eff}})(\rho_{\mathrm{eff}}')^2\rho^{-1} + \frac{2}{\rho}(0) &<0  & \nonumber \\
 f''(\rho_{\mathrm{eff}}) < 0& &
\label{eq:nudrho2}
\end{eqnarray}

Thus, using (\ref{eq:nudrho2}) in (\ref{eq:nutty}), we have the result that $\nu_R(P^*,t)$ is strictly concave w.r.t $t$.

QED \\

\section{Proof of proposition \ref{prop:optt}}
\label{sec:appproofoptt}
The equation that describes the optimal training time $t^*_s$ can be written as:

\begin{equation}
\label{eq:nuttymax}
(T-t^*_s) f'(\rho_{\mathrm{eff}})\rho_{\mathrm{eff}}'(t)_{t=S_d t^*_s} - f(\rho_{\mathrm{eff}})=0
\end{equation}

Now, let us study the optimal training time $t^*_s$ in the very high SNR regime, ie. when $\rho \to \infty$. (Note that $P \to \infty$ is equivalent to $\rho \to \infty$).

{ High SNR regime:} Applying the limit of $\rho \to \infty$ in (\ref{eq:nuttymax}), we get that

\begin{equation}
\label{eq:nuttyinf}
T_s-t^*_s = \ds{\lim_{\rho  \to \infty}} \frac{f(t\rho(1+t)^{-1})}{\rho f'(t\rho(1+t)^{-1})}
\end{equation}

We know from various works including \cite{veronica2} that  $\ds{\lim_{\rho  \to \infty}} f(t\rho(1+t)^{-1})=1$. Now let us consider $f'(\frac{t}{1+t} \rho)$.

For a MISO system we know from \cite{chis} that:
\begin{eqnarray}
f_{MISO}(\rho_{\mathrm{eff}}) &= & \frac{ { \it \gamma}\left( M, \frac{2^\xi-1}{\sqrt{\rho_{\mathrm{eff}}}} \right)  } {\Gamma(M)}
\label{eq:misoouts}
\end{eqnarray}where $\Gamma$ is the Gamma function and ${ \it \gamma}$ is the lower incomplete Gamma function. Now we can use the special property of the incomplete gamma function, $\ds{\lim_{x \to 0}} \frac{\gamma(M,x)}{x^M} = 1/M$  detailed in \cite{gamma} to determine $\ds{\lim_{\rho \to \infty} } \rho f(\rho_{\mathrm{eff}})' \propto \rho_{\mathrm{eff}}^{-M/2}$. Plugging this into (\ref{eq:nuttyinf}) we have:
\begin{equation}
\label{eq:nuttyinf2}
\ds{\lim_{\rho  \to \infty}} T_s-t^*_s =  \frac{1+t}{t}\rho_{\mathrm{eff}}^{M/2}
\end{equation}

Thus $\nu_R$ is optimized when $t_s \to - \infty$, but as $M \leq t_s <T_s$, we have $\ds{\lim_{\rho  \to \infty}} t^*_s=M$ for all MISO systems.

Now for any MIMO system in general, using the eigen value decomposition of $\Hm \Hm^H$ we have the eigenvalue decomposition $\log_2| { I_M} + \frac{\rho}{M}\Hm \Hm^H|=\log_2(\Pi_{i=1}^L(1+\lambda_i))$, where $L=\min(M,N)$ and $\lambda_i$ are the eigenvalues of $\Hm \Hm^H$. Applying the limit on $\rho$ and ignoring lower order terms we have
\begin{equation}
\ds{\lim_{\rho  \to \infty}} f(\rho_{\mathrm{eff}}) = \mathrm{Pr} \left[\Pi_{i=1}^L\lambda_i \geq \frac{2^\xi}{\rho_{\mathrm{eff}}^L}   \right].
\end{equation}
We can observe that the above expression is a cumulative distribution function of $\Pi_{i=1}^L\lambda_i$ and so it's derivative is simply the PDF of $\Pi_{i=1}^L\lambda_i$. For $\rho \to \infty$ we have $f'(\rho_{\mathrm{eff}})=\mathrm{Pr} [\Pi_{i=1}^L\lambda_i=\frac{2^\xi}{\rho_{\mathrm{eff}}^L} ]$ As we know that the in general, if the number of transmit antennas are the same, $\mathrm{Pr} [\lambda_{MISO} > x] \leq \mathrm{Pr} [\Pi_{i=1}^L\lambda_i > x]$ for any $x>0$ \cite{eigenrmt} . Thus $f_{MIMO}'(\rho_{\mathrm{eff}})<f_{MISO}'(\rho_{\mathrm{eff}})$, implying that for all MIMO systems $\ds{\lim_{\rho  \to \infty}} t^*_s=M$ from (\ref{eq:nuttyinf}) and (\ref{eq:nuttyinf2}).


%

%

\begin{IEEEbiography}[{\includegraphics[width=1in,height=1.25in,clip,keepaspectratio,bb=0 0 10 10]{varma.jpg}}]{Vineeth S Varma} was born in Tripunithura, India. He is currently doing his PhD at LSS under the supervision of Dr. Samson Lasaulce. He obtained his Bachelor’s in Physics with Honors from Chennai Mathematical Institute, India in 2008. He then proceeded to join the Erasmus Mundus course in Optics and obtained Masters in Science and Technology from Friedrich-Schiller-University of Jena in 2009 and Warsaw University of Technology in 2010. His areas of interest are Physics, Applied mathematics and Telecommunication. Since December 2010 he is working on his PhD aimed at finding the fundamental limits of energy on telecommunication and optimizing energy efficiency in networks. He is the recipient of the 2012 ACM VALUETOOLS best student paper award.
\end{IEEEbiography}

\begin{IEEEbiography}[{\includegraphics[width=1in,height=1.25in,clip,keepaspectratio, bb= 0 0 10 10]{lasau.jpg}}]{Samson Lasaulce} received his BSc and Agr\'egation degree in Physics from \'Ecole Normale Sup\'erieure (Cachan) and his MSc and PhD in Signal Processing from \'Ecole Nationale Sup\'erieure des T\'el\'ecommunications (Paris). He has been working with Motorola Labs for three years (1999, 2000, 2001) and with France T\'el\'ecom R\&D for two years (2002, 2003). Since 2004, he has joined the CNRS and Sup\'elec as a Senior Researcher. Since 2004, he is also Professor at \'Ecole Polytechnique. His broad interests lie in the areas of communications, information theory, signal processing with a special emphasis on game theory for communications. Samson Lasaulce is the recipient of the 2007 ACM VALUETOOLS, 2009 IEEE CROWNCOM, 2012 ACM VALUETOOLS best student paper award, and the 2011 IEEE NETGCOOP best paper award. He is an author of the book “Game Theory and Learning for Wireless Networks: Fundamentals and Applications”. He is currently an Associate Editor of the IEEE Transactions on Signal Processing.
\end{IEEEbiography}

\begin{IEEEbiography}[{\includegraphics[width=1in,height=1.25in,clip,keepaspectratio, bb= 0 0 10 10]{debba.jpg}}]{M\'erouane Debbah} entered the Ecole Normale Sup\'erieure de Cachan (France) in 1996 where he received his M.Sc and Ph.D. degrees respectively. He worked for Motorola Labs (Saclay, France) from 1999-2002 and the Vienna Research Center for Telecommunications (Vienna, Austria) until 2003. He then joined the Mobile Communications department of the Institut Eurecom (Sophia Antipolis, France) as an Assistant Professor until 2007. He is now a Full Professor at Supelec (Gif-sur-Yvette, France), holder of the Alcatel-Lucent Chair on Flexible Radio and a recipient of the ERC starting grant MORE (Advanced Mathematical Tools for Complex Network Engineering). His research interests are in information theory, signal processing and wireless communications. He is a senior area editor for IEEE Transactions on Signal Processing. M\'erouane Debbah is the recipient of the "Mario Boella" award in 2005, the 2007 General Symposium IEEE GLOBECOM best paper award, the Wi-Opt 2009 best paper award, the 2010 Newcom++ best paper award as well as the Valuetools 2007, Valuetools 2008, Valuetools 2012 and CrownCom2009 best student paper awards. He is a WWRF fellow. In 2011, he received the IEEE Glavieux Prize Award.
\end{IEEEbiography}
\begin{IEEEbiography}[{\includegraphics[width=1in,height=1.25in,clip,keepaspectratio, bb= 0 0 10 10]{elayo.jpg}}]{Salah Eddine Elayoubi} received his MS in telecommunications and networking from the National Polytechnic Institute at Toulouse, France, in 2001, and his Ph.D. and Habilitation in Computer science from the University of Paris VI in 2004 and 2009, respectively. Since then, he has been working in Orange Labs, the research and development division of France Telecom group. He is now a senior research engineer, involved in several EU-funded research projects. His research interests include radio resource management, modeling and performance evaluation of mobile networks.
\end{IEEEbiography}

\end{document}